\documentclass[aip,jmp,amsmath,amssymb,reprint]{revtex4-1}

\usepackage{color}
\usepackage{enumitem}
\usepackage{array}
\usepackage{float}
\usepackage{graphicx}
\usepackage{dcolumn}
\usepackage{bm}
\makeatletter

\newcommand{\Rmnum}[1]{\expandafter\@slowromancap\romannumeral #1@}
\makeatother

\begin{document}
\preprint{AIP/123-QED}

\title{Plasma instability of magnetically enhanced vacuum arc thruster}
\author{L. Chang}
\email{leichang@scu.edu.cn}
\affiliation{School of Aeronautics and Astronautics, Sichuan University, Chengdu 610065, China}
\author{T. P. Zhang}
\affiliation{National Key Laboratory of Science and Technology on Vacuum Technology and Physics, Lanzhou Institute of Physics, Lanzhou 730000, China}
\author{X. Y. Hu}
\affiliation{School of Aeronautics and Astronautics, Sichuan University, Chengdu 610065, China}
\author{X. M. Wu}
\affiliation{National Key Laboratory of Science and Technology on Vacuum Technology and Physics, Lanzhou Institute of Physics, Lanzhou 730000, China}
\author{X. F. Sun}
\affiliation{National Key Laboratory of Science and Technology on Vacuum Technology and Physics, Lanzhou Institute of Physics, Lanzhou 730000, China}
\date{\today}

\begin{abstract}
A two-fluid flowing plasma model is applied to describe the plasma rotation and resulted instability evolution in magnetically enhanced vacuum arc thruster (MEVAT). Typical experimental parameters are employed, including plasma density, equilibrium magnetic field, ion and electron temperatures, cathode materials, axial streaming velocity, and azimuthal rotation frequency. It is found that the growth rate of plasma instability increases with growing rotation frequency and field strength, and with descending electron temperature and atomic weight, for which the underlying physics are explained. The radial structure of density fluctuation is compared with that of equilibrium density gradient, and the radial locations of their peak magnitudes are very close, showing an evidence of resistive drift mode driven by density gradient. Temporal evolution of perturbed mass flow in the cross section of plasma column is also presented, which behaves in form of clockwise rotation (direction of electron diamagnetic drift) at edge and anti-clockwise rotation (direction of ion diamagnetic drift) in the core, separated by a mode transition layer from $n=0$ to $n=1$. This work, to our best knowledge, is the first treatment of plasma instability caused by rotation and axial flow in MEVAT, and is also of great practical interest for other electric thrusters where rotating plasma is concerned for long-time stable operation and propulsion efficiency optimization.  
\end{abstract}

\maketitle

\section{Introduction}\label{int}
\begin{figure}
\begin{center}$
\begin{array}{c}
\includegraphics[width=0.45\textwidth,angle=0]{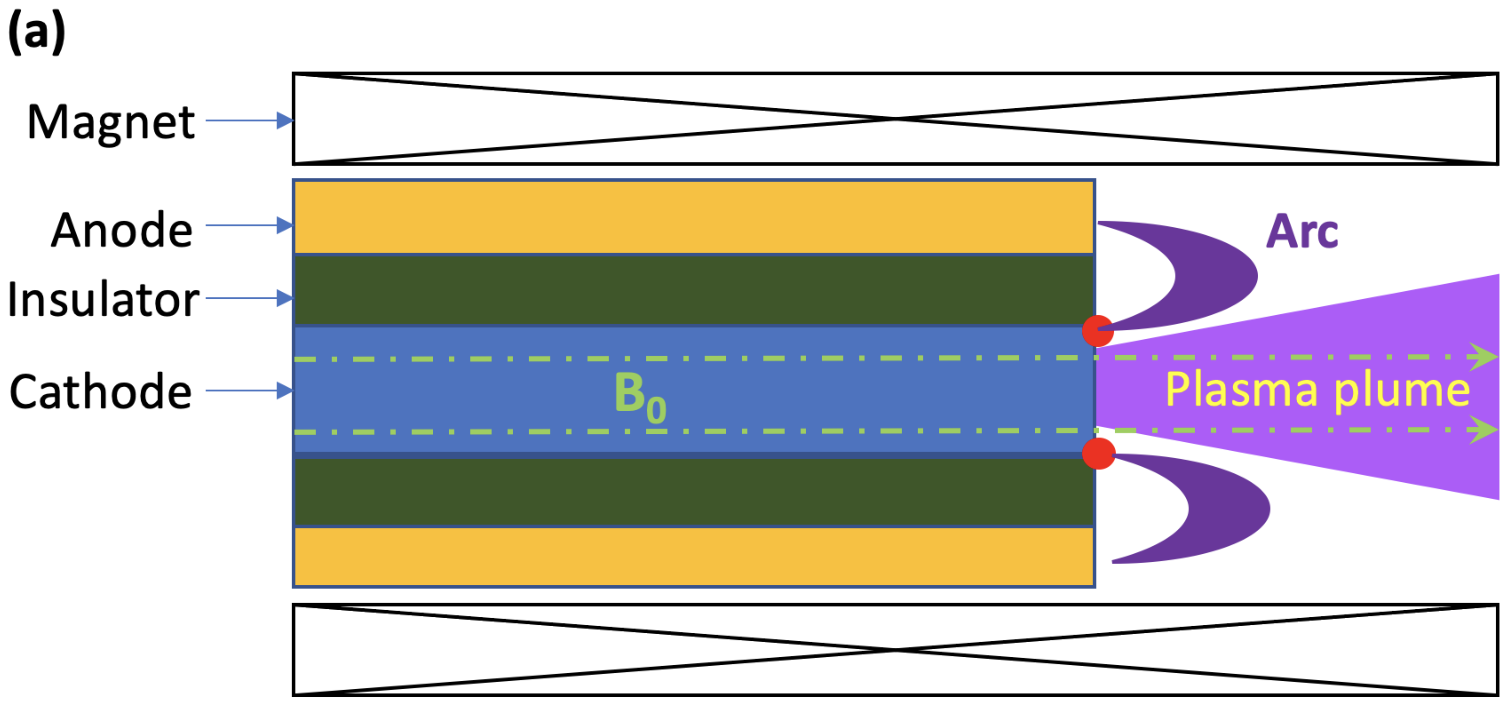}\\
\includegraphics[width=0.45\textwidth,angle=0]{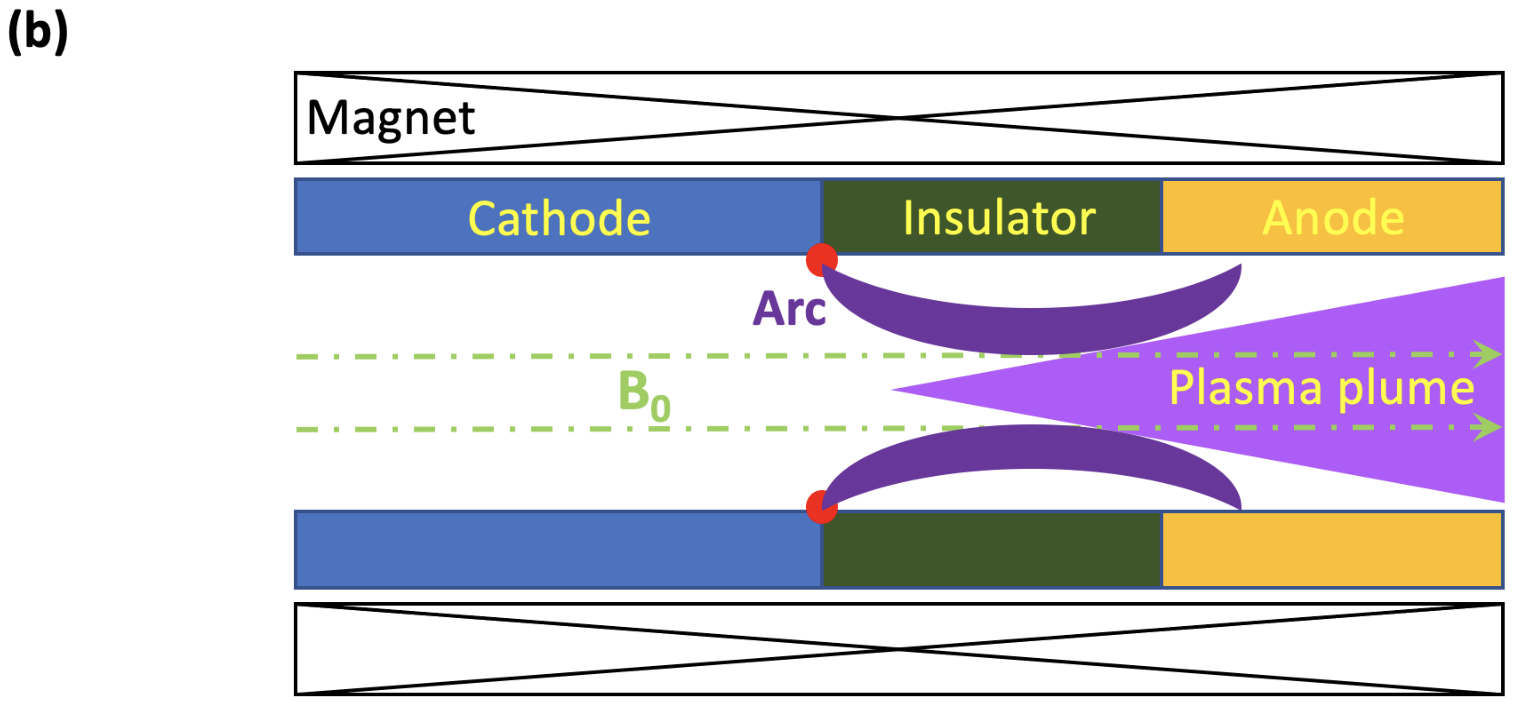}
\end{array}$
\end{center}\caption{Schematics of typical magnetically enhanced vacuum arc thruster (MEVAT): (a) coaxial-type, (b) ring-type.}
\label{fg1}
\end{figure}

Plasma propulsion is generally caused by jetting plasma in the opposite direction, according to Newton's third law, and accelerated either by electric force or magnetic force or both of them.\cite{Clark:1975aa, Martinez-Sanchez:1998aa, Charles:2009aa, Rafalskyi:2016aa, Mazouffre:2016aa, Levchenko:2018aa} External magnetic field is usually employed for efficient plasma generation and propulsion enhancement and control.\cite{Boswell:1970aa, Diaz:2003aa, Arefiev:2004ab, Charles:2006aa, Charles:2008aa, Lorello:2018aa} Due to the nonuniform configurations of equilibrium magnetic field, discharge area and plume, plasma rotation driven by Lorentz force commonly occurs in various electric thrusters.\cite{Corr:2009aa, Zhuang:2010aa, Fruchtman:2013aa, Aguirre:2017aa} A few analytical models have been developed or/and applied to describe this flowing phenomenon,\cite{Schein:2004aa, Chang:2011aa, Li:2013aa, Luskow:2018aa, Baranov:2018aa} but little attention was given to the resulted plasma instability which, however, can effect the propulsion efficiency, precise control, durable reliability and life time significantly.\cite{Choueiri:2001aa, Lafleur:2017aa} This paper considers an emerging plasma propulsion technology, namely magnetically enhanced vacuum arc thruster (MEVAT),\cite{Gilmour:1966aa, Gilmour:1967aa, Anders:1995aa, Keidar:1996aa, Keidar:2005aa} as an example and studies the instability evolution caused by plasma rotation and axial flow in detail. Schematics of the typical coaxial-type and ring-type MEVATs are shown in Fig.~\ref{fg1}, which illustrate the radial expansion of plasma plume across the confining field lines, leading to azimuthal rotation. Axial equilibrium field is usually employed to reduce the divergence of plasma plume and thereby increasing the propulsion efficiency, which is relatively low for unmagnetized vacuum arc thrusters, making the MEVAT a promising candidate to provide micro propulsion for small spacecrafts.\cite{Keidar:2005aa} The findings achieved here are also applicable to other types of electric thrusters with the involvement of external magnetic field, as long as the plasma rotation and flow are concerned for stable, efficient and safe operation. 

Specifically, a two-fluid flowing plasma model developed originally for interpreting wave oscillations in vacuum arc centrifuge,\cite{Hole:2002aa} which is a cylindrical, rapidly rotating, low temperature, and confined plasma column,\cite{Chang:2011aa} will be employed, together with a shooting numerical scheme for finding solutions. It will show that the instability strength peaks near the maximum of equilibrium density gradient, an evidence of resistive drift mode, and it becomes larger for higher rotation frequency, higher field strength, lower electron temperature and lower atomic weight. Temporal evolutions of perturbed mass flow in cross section will be also shown. The paper is organized as follows: Sec.~\ref{mdl} describes the theoretical model and steady-state and perturbed solutions, Sec.~\ref{cpt} presents the employed numerical scheme, dispersion relation, fluctuation structure and parameter dependence, and Sec.~\ref{cls} summarizes the whole paper and remarks the possible applications of this work to other electric thrusters.

\section{Two-fluid flowing plasma model}\label{mdl}
\subsection{Model assumptions}\label{ass}
The employed two-fluid flowing plasma model is based on the following assumptions:
{\setlength{\leftmargini}{12pt} 
\begin{enumerate}[topsep=0pt, partopsep=0pt]
\item Ions of different charge can be treated as a single species with average charge $Z$.
\item Plasma is quasi-neutral so that $n_e=Z n_i$.
\item Steady-state plasma is azimuthally symmetric and has no axial structure.
\item The effects of plasma fluctuation on external magnetic field is negligible.
\item Finite Larmor radius and viscosity effects are not considered.
\item Electron inertia can be neglected for the range of frequency considered. 
\item Ion and electron temperatures ($T_i$ and $T_e$) are uniform across the plasma column.
\item Steady-state ion density distribution is in form of $n_{0}=n_{i0}\exp[-(r/R)^2]$, with $n_{i0}$ the on-axis ion density and $R$ the characteristic radius where the density is $1/e$ of its on-axis value.
\item Steady-state velocities of ions and electrons are in forms of $\mathbf{v_i}=(0, \omega_i r, v_{iz})$ and $\mathbf{v_e}=[0, \omega_e(r) r, v_{ez}(r)]$, respectively, with $\omega_i$ the ion rigid rotor rotation frequency, $v_{iz}$ the ion uniform axial streaming velocity, $\omega_e(r)$ the electron rotation frequency, and $v_{ez}(r)$ the electron streaming velocity. 
\item Radial diffusion of both ions and electrons caused by electron-ion collision can be neglected.
\end{enumerate}
Moreover, length and time are normalized to $R$ and $1/\omega_{ic}$ respectively, with $\omega_{ic}=Z e B_z/m_i$ the ion cyclotron frequency, so that a normalized cylindrical coordinate system becomes $(x, \theta, \varsigma)=(r/R, \theta, z/R)$ and $\tau=\omega_{ic}t$, with $x$ and $\varsigma$ the normalized radial and axial positions, respectively. 

\subsection{Governing equations}\label{eqs}
The model consists of the momentum and continuity equations for ion and electron fluids:
\begin{equation}\label{eq1}
\frac{\partial \mathbf{u_i}}{\partial \tau}+(\mathbf{u_i} \cdot \mathbf{\nabla}) \mathbf{u_i} = -\psi(\rm{Z} \mathbf{\nabla} \chi+\lambda \mathbf{\nabla} l_i)+\mathbf{u_i} \times \mathbf{\hat{\varsigma}}+\delta \rm{n_s} \mathbf{\tilde{\xi}} \cdot (\mathbf{u_e}-\mathbf{u_i}),
\end{equation}
\begin{equation}\label{eq2}
\psi Z(-\mathbf{\nabla} l_i+\mathbf{\nabla} \chi)-\mathbf{u_e} \times \mathbf{\hat{\varsigma}}+\delta \rm{n_s} \mathbf{\tilde{\xi}} \cdot (\mathbf{u_i}-\mathbf{u_e})=\rm{0},
\end{equation}
\begin{equation}\label{eq3}
-\frac{\partial l_i}{\partial \tau}=\mathbf{\nabla} \cdot \mathbf{u_i}+\mathbf{u_i} \cdot \mathbf{\nabla} \rm{l_i},
\end{equation}
\begin{equation}\label{eq4}
-\frac{\partial l_i}{\partial \tau}=\mathbf{\nabla} \cdot \mathbf{u_e}+\mathbf{u_e} \cdot \mathbf{\nabla} \rm{l_i},
\end{equation}
\\with terms defined as: 
\[\mathbf{u_i}=\frac{\mathbf{v_i}}{\omega_{ic} \rm{R}}=(\rm{x \varphi_i}, \rm{x \Omega_i}, \rm{u_{i\varsigma}}), \mathbf{u_e}=\frac{\mathbf{v_e}}{\omega_{ic}\rm{R}}=(\rm{x \varphi_e}, \rm{x \Omega_e}, \rm{u_{e\varsigma}}),\]
\[\lambda=\frac{T_i}{T_e}, \psi=\frac{k_B T_e}{m_i \omega_{ic}^2 R^2}, \chi=\frac{e \phi}{k_B T_e}, \mathbf{\tilde{\xi}}=diag(\xi_\bot, \xi_\bot, 1),\]
\[l_i=ln \frac{n_i}{n_{i0}}, n_s=\frac{\nu_{ei}}{\nu_{ei0}}, \delta=\frac{e Z n_{i0}}{B_z} \frac{\eta_L}{\gamma_E}.\]
The subscript $i$ and $e$ refer to ion and electron parameters respectively, $\varphi$ represents the normalized radial velocity divided by $x$, $\Omega$ stands for the normalized rotation frequency, $u_{\varsigma}$ shows the normalized axial velocity, $\lambda$ labels the ratio between ion and electron temperatures, $\psi$ behaves a convenient constant which for $\lambda=1$ becomes the square of the normalized ion thermal velocity, $\chi$ gives a normalized electric potential $\phi$, $l_i$  expresses the logarithm of the ratio of ion density $n_i$ to its on-axis value $n_{i0}$, and $n_s$ means the ratio of electron-ion collision frequency $\nu_{ei}$ to its on-axis value $\nu_{ei0}$. Moreover, $\delta$ phrases the normalized resistivity parallel to external magnetic field, with $\eta_L$ the electrical resistivity of Lorentz gas and $\gamma_E$ the ratio of the conductivity of charge state $Z$ to that in Lorentz gas.\cite{Spitzer:1962aa}

\subsection{Steady-state solution}\label{std}
For constant magnetic field of ${\mathbf B}=(0, 0, B_z)$, the steady-state solution is
\begin{equation}\label{eq5}
\chi_0(x)=\chi_c+[\frac{\Omega_{i0}}{2\psi Z}(1+\Omega_{i0})+\frac{\lambda}{Z}]x^2, 
\end{equation}
with
\begin{equation}\label{eq6}
\Omega_{e0}=\Omega_{i0}(1+\Omega_{i0})+2 \psi (\lambda+Z).
\end{equation}
Here, $\chi_c$ is an arbitrary reference potential. Because the axial current in this model is unconstrained, which is consistent with MEVAT boundary conditions, it can be arbitrarily set to zero ($u_{i\varsigma0}=u_{e\varsigma0}$).

\subsection{Perturbed solution}\label{ptb}
To solve the perturbed solution, a linear perturbation treatment with plasma parameters $\zeta$ is considered, namely
\begin{equation}\label{eq7}
\zeta (\tau, x, \theta, \varsigma)=\zeta_0(x)+\varepsilon \zeta_1(x)e^{i(m \theta+k_\varsigma \mathbf{\varsigma}-\omega \tau)}.
\end{equation}
Here, $\varepsilon$ is the perturbation parameter, $m$ is the azimuthal mode number, $k_\varsigma$ is the axial wave number and $\omega$ is the angular frequency. To first order of $\varepsilon$,  Eqs.~(\ref{eq1})-(\ref{eq4}) can be reduced to
\begin{equation}\label{eq8}
\left(
\begin{array}{clcr}
\psi (l_{i1}'(y)-X_1'(y))\\
y \varphi_{i1}'(y)\\
y \varphi_{e1}'(y)-i m \Psi X_1'(y)\\
\Psi \delta \xi_\bot e^{-y} X_1'(y)\\
0
\end{array}
\right)
=
\tilde{\mathbf{A}}
\left(
\begin{array}{clcr}
l_{i1}(y)\\
X_1(y)\\
\varphi_{i1}(y)\\
\varphi_{e1}(y)\\
u_{e\varsigma}(y)
\end{array}
\right)
\end{equation}
with $\tilde{\mathbf{A}}$ the matrix
\begin{widetext}
\begin{equation}\label{eq9}
\begin{array}{lll}
\tilde{\mathbf{A}} & = &
\left(
\begin{array}{ccccc}
\frac{m \Psi C}{2 \varpi y}                                                 & 0                   & \frac{i \varpi}{2}-\frac{i C^2}{2 \varpi} & \frac{i C}{2 \varpi} & 0\\
\frac{i}{2}(\varpi-\frac{\Psi}{\varpi}(\frac{m^2}{y}+k_\varsigma^2)) & 0                   & -1+y-\frac{m C}{2 \varpi}                       & \frac{m}{2 \varpi}   & 0\\
\frac{i}{2}(\varpi-m \Omega_{i0}^2-2m \Psi)                                & 0                   & 0                                                     & -1+y                       & -\frac{i k_\varsigma}{2}\\
0                                                                                & \frac{i m \Psi}{2y} & 0                                                     & -\frac{1}{2}               & 0\\
0                                                                                & -i k_\varsigma \Psi & 0                                                     & 0                          & 0\\
\end{array}
\right)\\
&+&\delta
\left(
\begin{array}{ccccc}
0                                                                                & 0 & -\frac{\xi_\bot e^{-y}}{2}                   & \frac{\xi_\bot e^{-y}}{2}                   & 0\\
0                                                                                & 0 & 0                                            & 0                                           & 0\\
0                                                                                & 0 & -\frac{i m \xi_\bot e^{-y}}{2}               & \frac{i m \xi_\bot e^{-y}}{2}               & 0\\
\xi_\bot e^{-y}(-\frac{m \Psi}{2 \varpi y}+\frac{\Omega_{i0}^2}{2}+\Psi)   & 0 & \frac{i C \xi_\bot e^{-y}}{2 \varpi}   & -\frac{i \xi_\bot e^{-y}}{2 \varpi}   & 0\\
\frac{e^{-y}k_\varsigma \Psi}{\varpi}                                      & 0 & 0                                            & 0                                           & -e^{-y}
\end{array}
\right).
\end{array}
\end{equation}
\end{widetext}
Here, $\varpi=\omega-m \Omega_{i0}-k_\varsigma u_{\varsigma 0}$ is the frequency in the frame of ion fluid, $C=1+2\Omega_{i0}$, $\Psi=(\lambda+Z)\psi$, $y=x^2$ and a new dependent variable $X_1(y)$ has been introduced
\begin{equation}\label{eq10}
X_1(y)=\frac{Z}{Z+\lambda}[l_{i1}(y)-\chi_1(y)]. 
\end{equation}
For large axial wavelength modes of the resistive plasma column, i. e. $k_\varsigma^2\leq\delta$, Eq.~(\ref{eq8}) turns out to be a second order differential equation
\begin{equation}\label{eq11}
(\frac{\varpi^2-C^2}{\varpi\Psi})L(N_c)[g_1(y)]=0,
\end{equation}
where
\[L(N_c)=y \frac{\partial^2}{\partial y^2}+(1-y)\frac{\partial}{\partial y}+(\frac{N_c}{2}-\frac{m^2}{4y}),\]
\[N_c=\frac{(\varpi^2-C^2)(m+\frac{i}{2}f(y))}{\varpi-m \Omega_{i0}^2+i \Psi f(y)}+\frac{m C}{\varpi},\]
and $f(y)=F^2 e^{y}$ with the normalized axial wave number $F=k_\varsigma/\sqrt{\delta}$. For odd $m$ modes, the boundary conditions are $g_1(0)=0$ and $g_1(Y)=0$ with the infinite radius $Y$ representing the edge of plasma column. For even $m$, these conditions become $g_1'(0)=0$ and $g_1(Y)=0$. We only consider unstable solutions for which $\varpi^i>0$. The solutions $\varpi=\pm C$ are stable and thereby discarded.  

\section{Computed wave physics analysis}\label{cpt}
\subsection{Numerical scheme and conditions}\label{num}
To solve Eq.~(\ref{eq11}) numerically for perturbed solutions, we make use of a shooting method, as did by Hole et al.\cite{Hole:2002aa} For $m=1$ mode, the boundary conditions are $g_1(0)=g_1(Y)=0$. The gradient at edge $g_1'(Y)$ is arbitrary because the differential equation is homogeneous. As a result, we set $g_1'(Y)=1$. For given $F$, a trial $\varpi$ is first chosen and then the solution is matched from edge to core. We adjust the complex frequency $\varpi$ until the on-axis boundary condition is satisfied. The procedure starts from $F~=~0$, for which an analytical solution for $\varpi$ can be found from
\begin{equation}\label{eq12}
N_c=\frac{m(\varpi^2-C^2)}{\varpi-m \Omega_{i0}^2}+\frac{m C}{\varpi}=2 n+|m|
\end{equation}
with $n$ the number of radial nodes in the plasma column. This numerical scheme has been benchmarked by previous studies.\cite{Hole:2002aa, Chang:2011aa}

For the computational parameters, we refer to existing devices and choose published experimental data. Table~\ref{tab1} shows the typical parameters of MEVAT employed for the present computation, together with those of a plasma centrifuge (PCEN\cite{Hole:2002aa, Dallaqua:1998aa}) for comparison. Although MEVAT has much higher electron and ion temperatures and lower rotation rate, the two-fluid model originally developed for PCEN can still well describe the flowing plasma in MEVAT. The plasma density of $n_{i0}=5\times 10^{19}~\rm{m^{-3}}$ is close to the measurement by Keidar et al.\cite{Keidar:2014aa} Trial electron and ion temperatures are $100~\rm{eV}$ and $50~\rm{eV}$, respectively, for current in order of $1~\rm{kA}$\cite{Zverev:1998aa} and with reference to a previous experiment.\cite{Keidar:2014aa} The strength of external magnetic field $B_z=0.1~\rm{T}$ is typical for various MEVATs.\cite{Keidar:2005aa, Baranov:2018aa} The axial stream velocity of ion fluid is set to $1.3\times 10^4~\rm{m~s^{-1}}$ according to the experiment using titanium (Ti) cathode,\cite{Kutzner:1992aa} which lies inside the typical range of ion velocity $10-50~\rm{km~s^{-1}}$,\cite{Dethlefsen:1968aa, Qi:2001aa, Schein:2002aa, Schein:2002ab, Schein:2003aa} and meanwhile the ion species of Ti is chosen. The rotation frequency is assumed to be half the ion cyclotron frequency, which is close to that of plasma centrifuge\cite{Dallaqua:1998aa, Hole:2002aa} to see the effect of rotation on instability evolution more clearly, and rigid rotation is a reasonable assumption according to various jetting plasma devices.\cite{Prasad:1987aa, Dallaqua:1998aa, Chang:2011aa} Moreover, charge number of $Z=2$ is considered throughout the paper as a common value for different cathode materials chosen here (see Sec.~\ref{prm}), and characteristic radius of $R=0.03~\rm{m}$ is made use of referring to previous studies.\cite{Keidar:2005aa, Luskow:2018aa}
\begin{table}
\caption{\label{tab1}Typical parameters of MEVAT and PCEN (plasma centrifuge).}
\begin{ruledtabular}
\begin{tabular}{lll}
Parameter      & MEVAT\cite{Keidar:2014aa, Zverev:1998aa, Keidar:2005aa, Baranov:2018aa, Kutzner:1992aa, Dallaqua:1998aa, Hole:2002aa,  Luskow:2018aa}  & PCEN\cite{Dallaqua:1998aa,Hole:2002aa} \\
\hline
$n_{i0}$ (on axis)                                         & $5\times 10^{19}\ \rm{m^{-3}}$   & $5.2\times10^{19}\ \rm{m^{-3}}$\\
$T_e$                                                           & $100$ eV                           & $2.9$ eV\\
$T_i$                                                            & $50$ eV                           & $2.9$ eV\\
$m_i$                                                           & $47.87$ amu (Ti)                           & $24.31$ amu (Mg)\\
$B_z$                                                           & $0.1$ T                         & $0.05$ T\\
$Z$                                                               & $2.0$                              & $1.5$\\
$V_{z0}$                                                      & $1.3\times 10^{4}\ \rm{m~s^{-1}}$               & $10^4\ \rm{m~s^{-1}}$\\
$\omega_0$                                                 & $201\ \rm{krad~s^{-1}}$           & $184\ \rm{krad~s^{-1}}$\\
$\omega_{ic}= \frac{B_z e Z}{m_i}$                              & $402\ \rm{krad~s^{-1}}$           & $295\ \rm{krad~s^{-1}}$\\
$\Omega_{i0}=\frac{\omega_0}{\omega_{ic}}$                      & $0.5$                            & $0.59$\\
$\Psi=(\frac{T_i}{T_e}+Z)\frac{k_B T_e}{m_i \omega_{ic}^2 R^2}$ & $3.46$                             & $1.6$ \\
$\delta=\frac{e Z n_{i0}}{B_z} \frac{\eta_L}{\gamma_E}$         & $7.8\times 10^{-5}$                           & $0.03$\\
$R$(characteristic radius)                                      & $3$ cm                             & $1.43$ cm
\end{tabular}
\end{ruledtabular}
\end{table}
The radial density profile is shown in Fig.~\ref{fg2}, a typical plasma distribution near the exit of MEVAT.\cite{Keidar:1996aa, Keidar:2005aa} We shall compute the dispersion relation and instability physics based on these experimental data in the following sections.  
\begin{figure}
\begin{center}
\includegraphics[width=0.45\textwidth,angle=0]{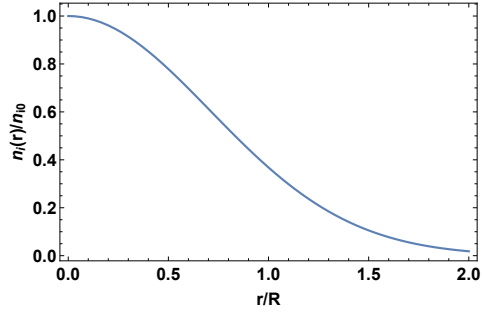}\\
\end{center}\caption{Typical radial profile of plasma density near the exit of MEVAT.\cite{Keidar:1996aa, Keidar:2005aa}}
\label{fg2}
\end{figure}

\subsection{Dispersion relation}\label{dsp}
Before presenting the obtained dispersion curve, we should check the eigenfunction associated with the computation. A typical radial variation of solved eigenfunction for $F=0$ is shown in Fig.~\ref{fg3}, which clearly satisfies the boundary conditions of $g_1(0)=g_1(Y)=0$ for $m=1$ mode. Moreover, it shows a radial mode of $n=1$ with visible node located near $y=2$. 
\begin{figure}
\begin{center}
\includegraphics[width=0.45\textwidth,angle=0]{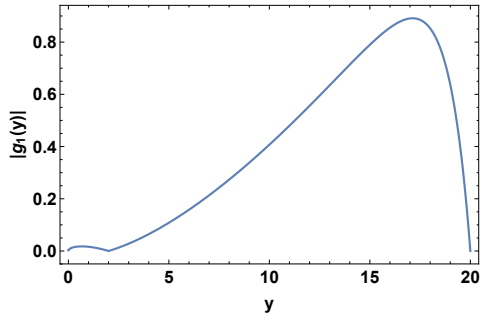}\\
\end{center}\caption{Radial variation of solved eigenfunction for $F=0$.}
\label{fg3}
\end{figure}
The computed dispersion curves in the range of $F=0-1$ are illustrated in Fig.~\ref{fg4}, compared with those for PCEN. It can be seen that the normalized growth rate of MEVAT peaks on axis and decreases monotonously with $F$, whereas the normalized growth rate of PCEN peaks off axis ($F=0.3$) and exhibits a mode crossing near $F=0.55$, which is caused by centrifugal instability. Relatively, the instability level (proportional to growth rate) is much more lower on MEVAT than that on PCEN for $F>0.3$, due to significantly smaller resistivity along the external magnetic field (it will show later that the instability is a resistive drift mode). Further, the peak growth rate of $\varpi^i_{\rm{max}}=0.324$ corresponds to $\varpi^r=-0.205$ for MEVAT, implying that the instability frequency is smaller than the sum of plasma rotation frequency and axial velocity, because $\varpi=\omega-m \Omega_{i0}-k_\varsigma u_{\varsigma 0}$ is the frequency in the frame of ion fluid, and it propagates in the $-\theta$ direction (same to the direction of ion diamagnetic drift); whereas, the peak growth rate of $\varpi^i_{\rm{max}}=0.39$ corresponds to $\varpi^r=-0.007$ for PCEN, stating that the instability is near stationary in the frame of ion fluid. Overall, the normalized frequency is lower on MEVAT for $F>0.4$, which may be attributed to lower rotation frequency and ion temperature as revealed in a previous study.\cite{Chang:2011aa}    
\begin{figure}
\begin{center}$
\begin{array}{c}
\includegraphics[width=0.45\textwidth,angle=0]{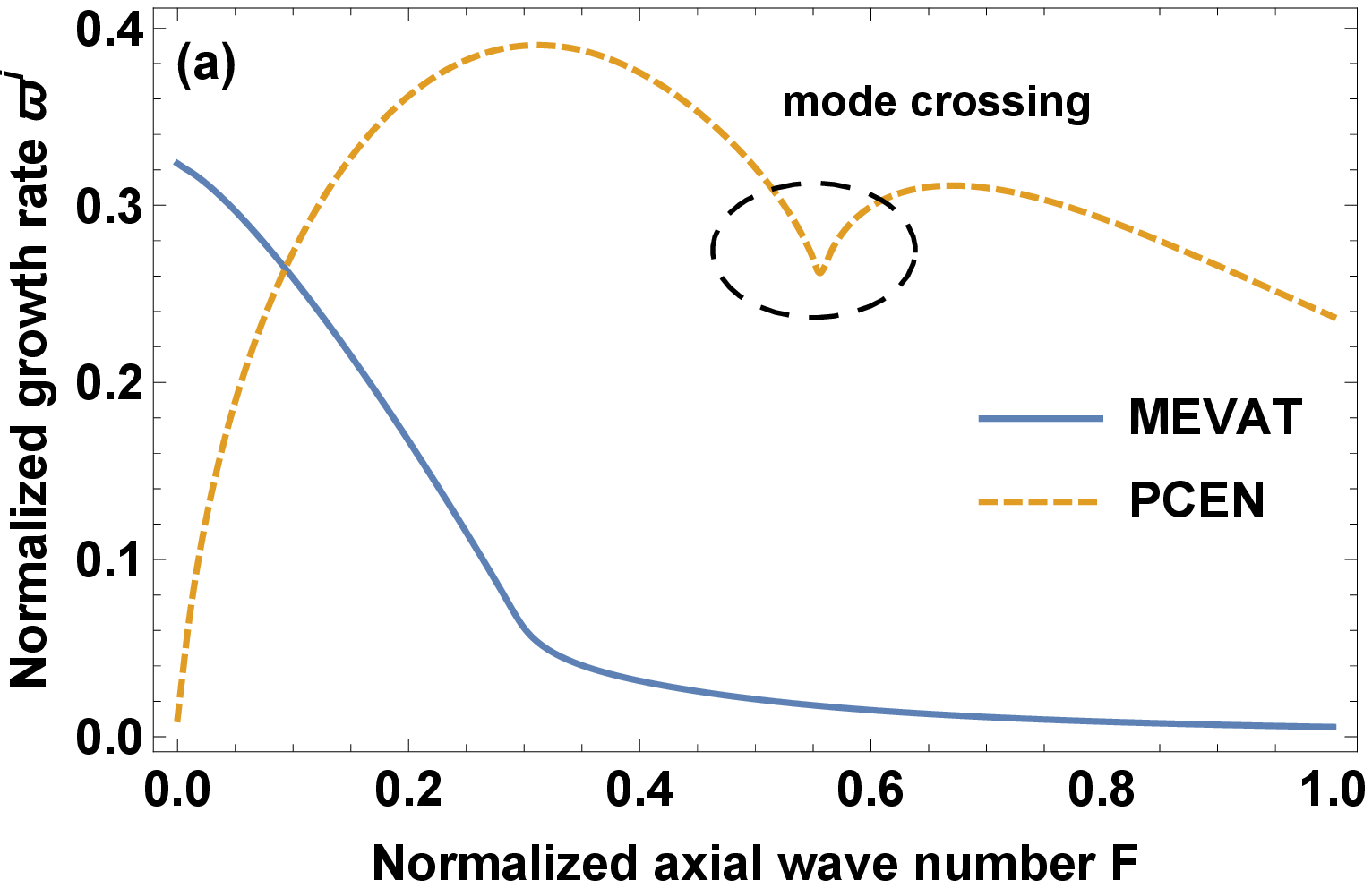}\\
\hspace{-0.18 cm}\includegraphics[width=0.462\textwidth,angle=0]{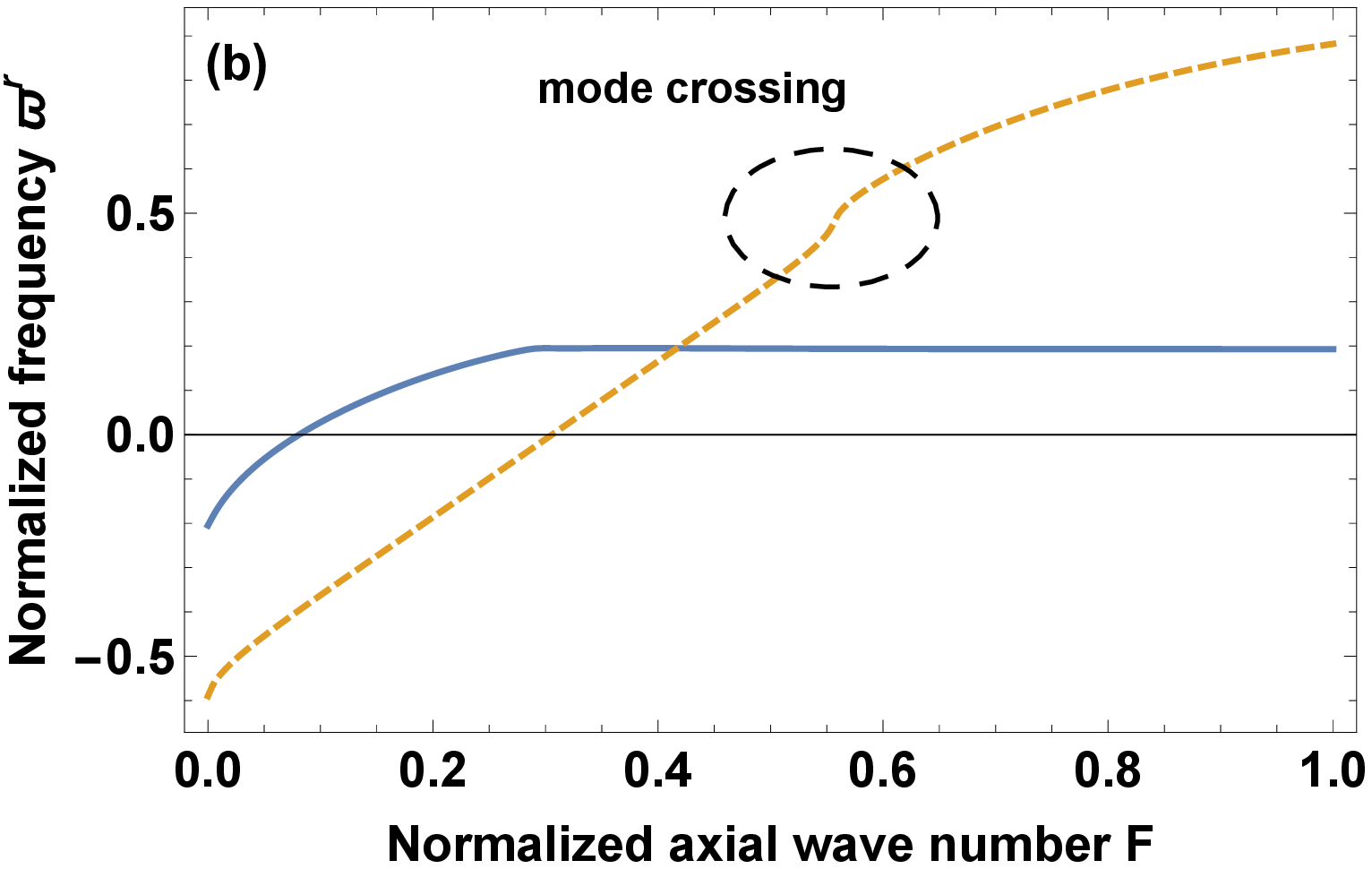}
\end{array}$
\end{center}\caption{Dispersion curves for MEVAT (solid line) and PCEN (dashed line), generated based on the conditions shown in Table~\Rmnum{1}: (a) normalized growth rate $\varpi^i$; (b) normalized frequency $\varpi^r$. (vs normalized axial wavenumber $F~=~k_\varsigma/\sqrt{\delta}$).}
\label{fg4}
\end{figure}

\subsection{Fluctuation structure}\label{flc}
To show a cross sectional view of the mode structure, the vector field of linearly perturbed mass flow was calculated through $m_i (n_{i1} \mathbf{u_{i0}}~+~n_{i0} \mathbf{u_{i1}})$. We computed the perturbed velocity components ${\mathbf u_{i1}}~=~(\rm{x \varphi_{i1}},~\rm{x \Omega_{i1}},~\rm{u_{i\varsigma1}})$,${\mathbf u_{e1}}~=~(\rm{x \varphi_{e1}},~\rm{x \Omega_{e1}},~\rm{u_{e\varsigma1}})$ and perturbed density $n_{i1}$ from the solution of $g_1(y)$ and following equations:\cite{Chang:2011aa}
\begin{equation}
l_{i1}(y)=\frac{-g_1(y)}{(1+\frac{i}{\Psi}(\frac{m \Omega_{i0}^2+2m \Psi-\varpi}{f(y)-2 i m}))},
\end{equation}
\begin{equation}
\chi_1(y)=-\frac{\lambda}{Z}l_{i1}(y)-(1+\frac{\lambda}{Z})g_1(y),
\end{equation}
\begin{equation}
\setlength{\extrarowheight}{0.3cm}
\begin{array}{rcl}
\varphi_{i1}(y) & = & \frac{2\varpi}{i(\varpi^2-C^2)-\varpi \delta \xi_\bot e^{-y}}[\Psi(l_{i1}'(y)-X_1'(y))\\
&&-\frac{m \Psi C}{2\varpi y}l_{i1}(y)-(\frac{i C}{2\varpi}+\frac{\delta \xi_\bot e^{-y}}{2})\varphi_{e1}(y)],
\end{array}
\end{equation}
\begin{equation}
\setlength{\extrarowheight}{0.3cm}
\begin{array}{rcl}
\varphi_{e1}(y) & = &\frac{2\varpi}{\varpi+i \delta \xi_\bot e^{-y}}[\frac{i m \Psi}{2y}X_1(y)-\Psi \delta \xi_\bot e^{-y} X_1'(y)\\
&&+\delta \xi_\bot e^{-y}(-\frac{m \Psi}{2 \varpi y}+\frac{\Omega_{i0}^2}{2}+\Psi)l_{i1}(y)\\
&&+\frac{i C \delta \xi_\bot e^{-y}}{2\varpi}\varphi_{i1}(y)],
\end{array}
\end{equation}
\begin{equation}
\Omega_{i1}(y)=\frac{1}{\varpi}[m \Psi \frac{l_{i1}(y)}{y}+i(\varphi_{e1}(y)-C\varphi_{i1}(y))], 
\end{equation}
\begin{equation}
u_{iz1}(y)=\frac{\sqrt{\delta} F \Psi}{\varpi}l_{i1}(y).
\end{equation}
Figure~\ref{fg5} displays the radial profiles of perturbed density $n_{i1}(r)$ and equilibrium density gradient $|n'_{i0}(r)|$. We can see that the perturbed density peaks ($6.89\times 10^{15}~\rm{m^{-3}}$) at $r=1.68$~cm, which is very close to the radial location $r=2.12$~cm of maximum density gradient ($1.43\times 10^{21}~\rm{m^{-3}}$) in equilibrium state. This suggests that the observed instability shown in Fig.~\ref{fg4} may be a resistive drift mode, which is driven by plasma density gradient. 
\begin{figure}
\begin{center}
\includegraphics[width=0.47\textwidth,angle=0]{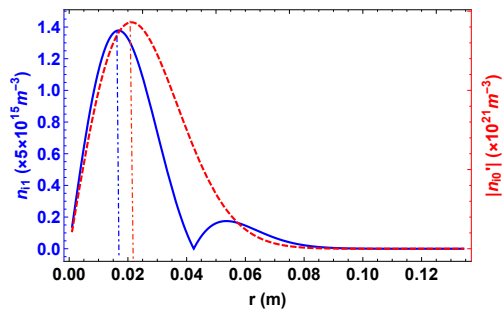}
\end{center}
\caption{Radial variations of perturbed density $n_{i1}(r)$ (solid) and equilibrium density gradient $|n'_{i0}(r)|$ (dashed line).}
\label{fg5}
\end{figure}
Temporal evolution of the cross sectional mass flow is given in Fig.~\ref{fg6}, for a period of $t=1$~s. The time dependence was achieved by multiplying $\exp[i(2\pi t)]$ with $m_i (n_{i1} \mathbf{u_{i0}}~+~n_{i0} \mathbf{u_{i1}})$. Here, the external magnetic field and $z$ point into the page, and coordinates $x$ and $y$ label the cross section of plasma column, namely $x=r\cos \theta$ and $y=r\sin \theta$. We can see that there exists a radial layer inside which the mass flow rotates in the anti-clockwise direction (same to the direction of ion diamagnetic drift), while it rotates in the clockwise direction (same to the direction of electron diamagnetic drift) outside, indicating a circularly sheared flow near the layer. The radial location of this layer is around the mode transition radius of $r=4.24$~cm shown in Fig.~\ref{fg5}, from $n=0$ mode to $n=1$ mode. This new pattern of rotation is different from the symmetric rotation observed before,\cite{Chang:2011aa}, and may be attributed to the large difference between electron and ion temperatures and very low normalized resistivity along the equilibrium magnetic field. Moreover, the mass flow is largest around the peak density gradient at $r=2.12$~cm, and drops to zero when approaching to the core and edge of plasma column, consistent with the boundary conditions of perturbed density (Fig.~\ref{fg5}). Although only linear oscillatory response is considered here, the rotation may be damped by a similar nonlinear flow pattern as claimed before.\cite{Chang:2011aa}
\begin{figure*}
\begin{center}$
\begin{array}{cccc}
\includegraphics[width=0.24\textwidth,angle=0]{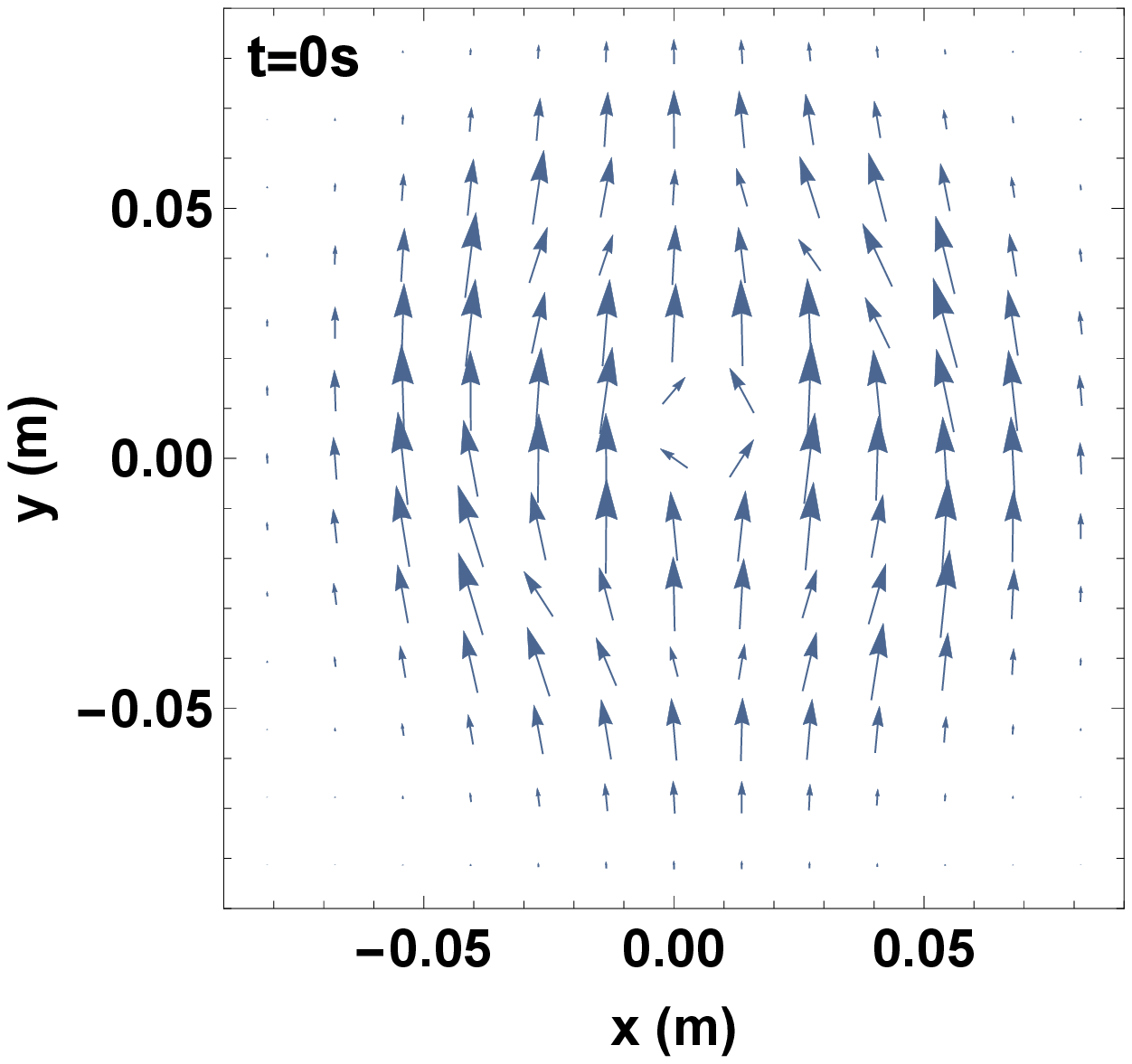}&\includegraphics[width=0.24\textwidth,angle=0]{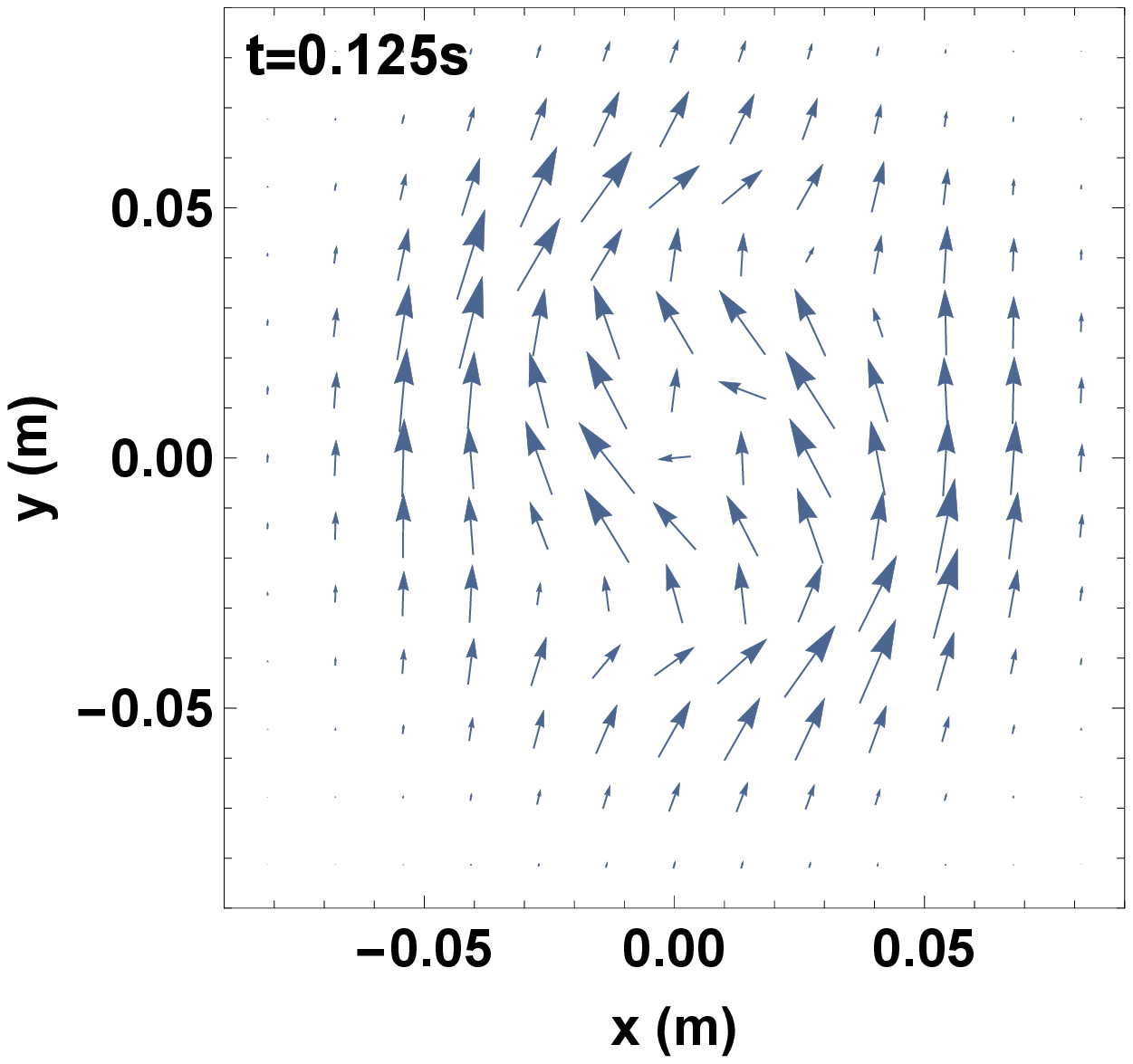}&\includegraphics[width=0.24\textwidth,angle=0]{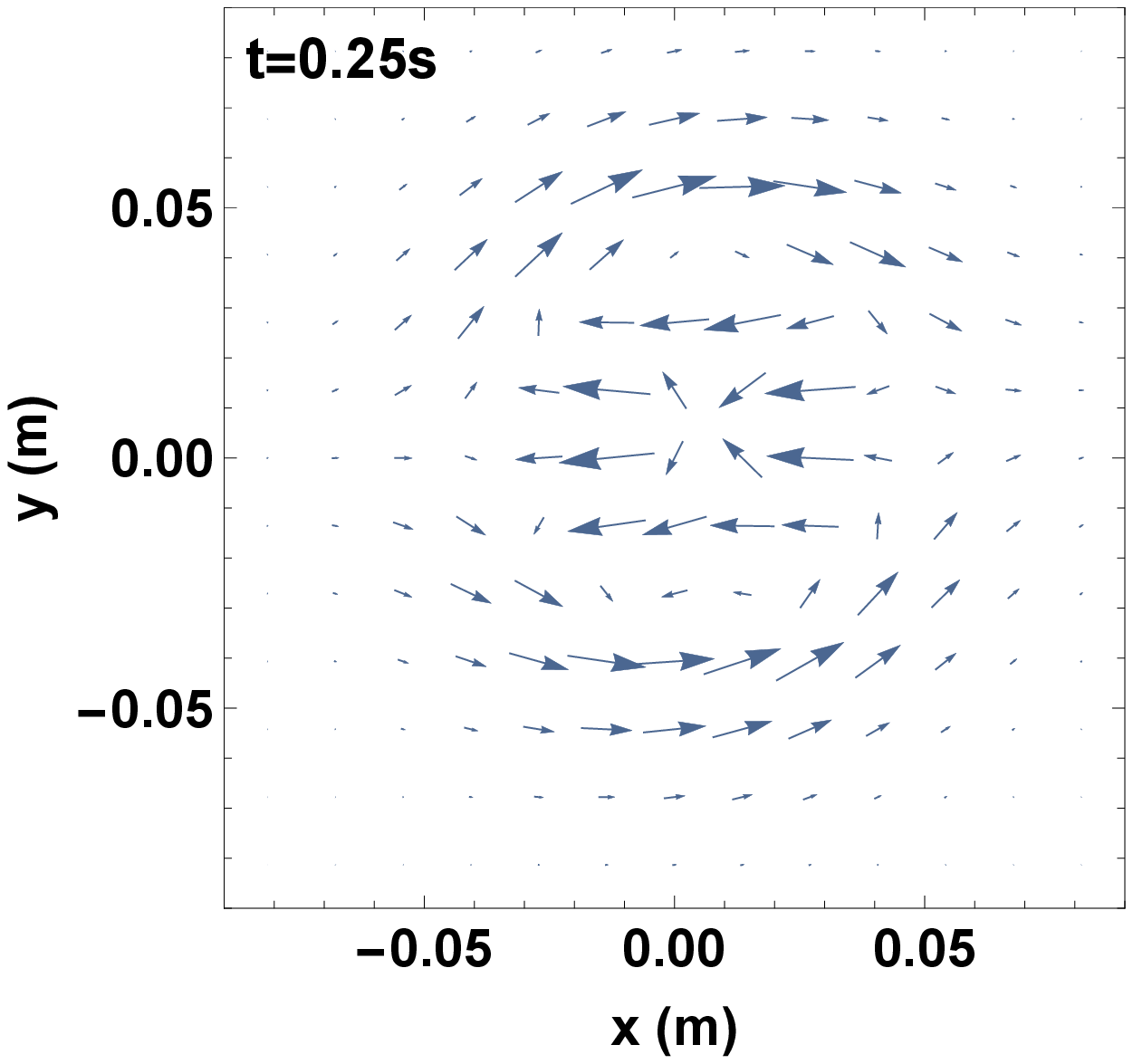}&\includegraphics[width=0.24\textwidth,angle=0]{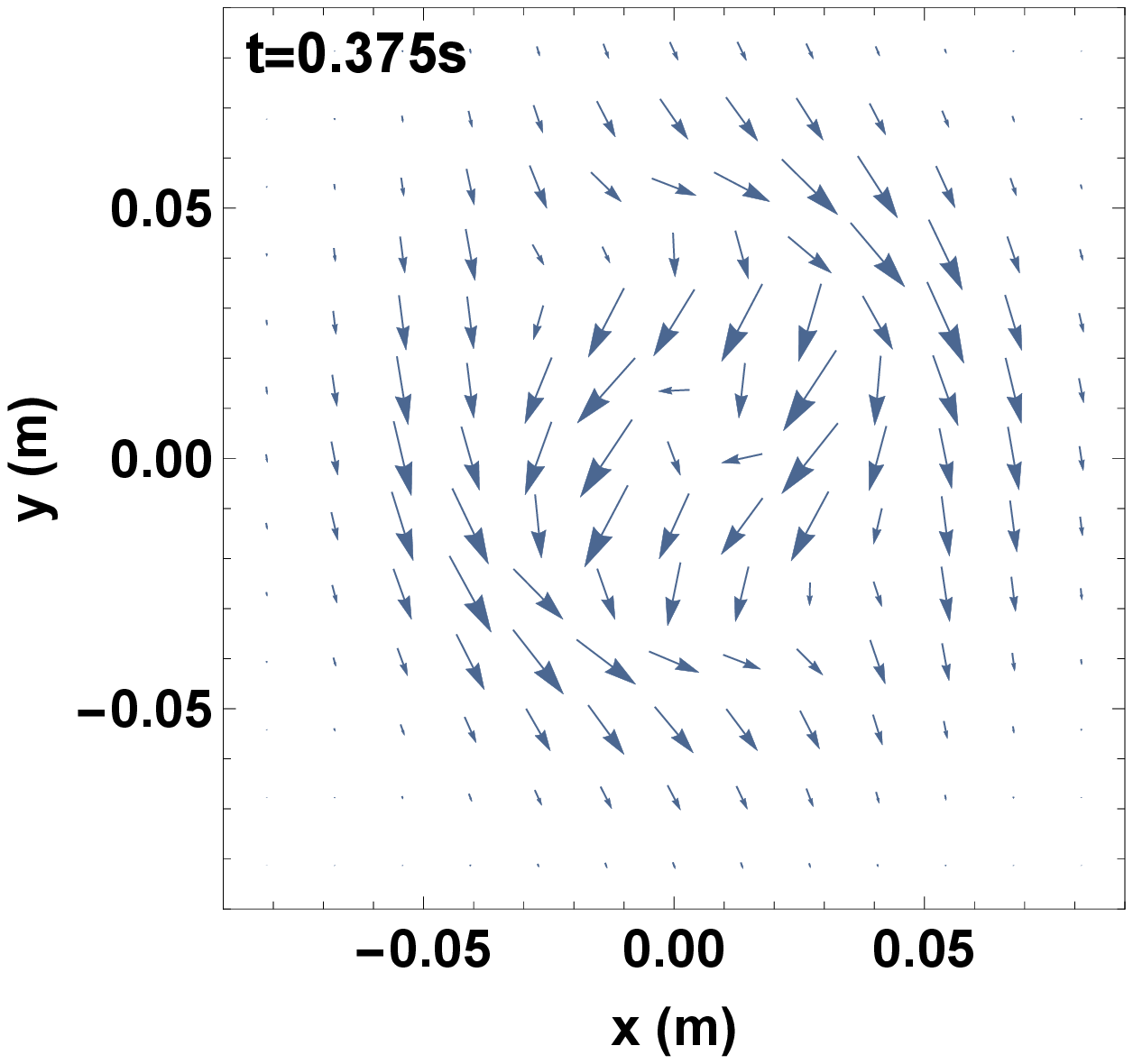}\\
\includegraphics[width=0.24\textwidth,angle=0]{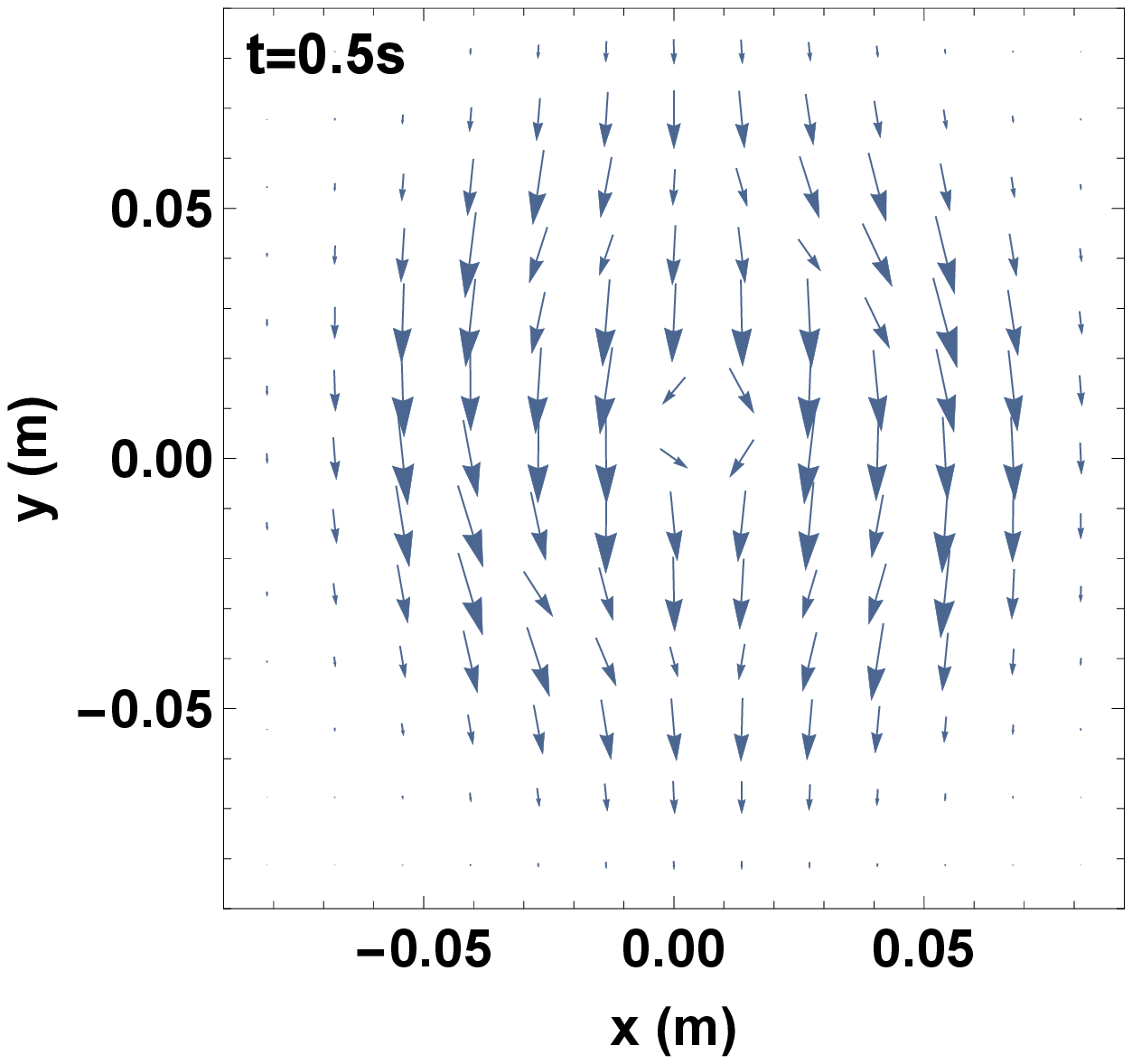}&\includegraphics[width=0.24\textwidth,angle=0]{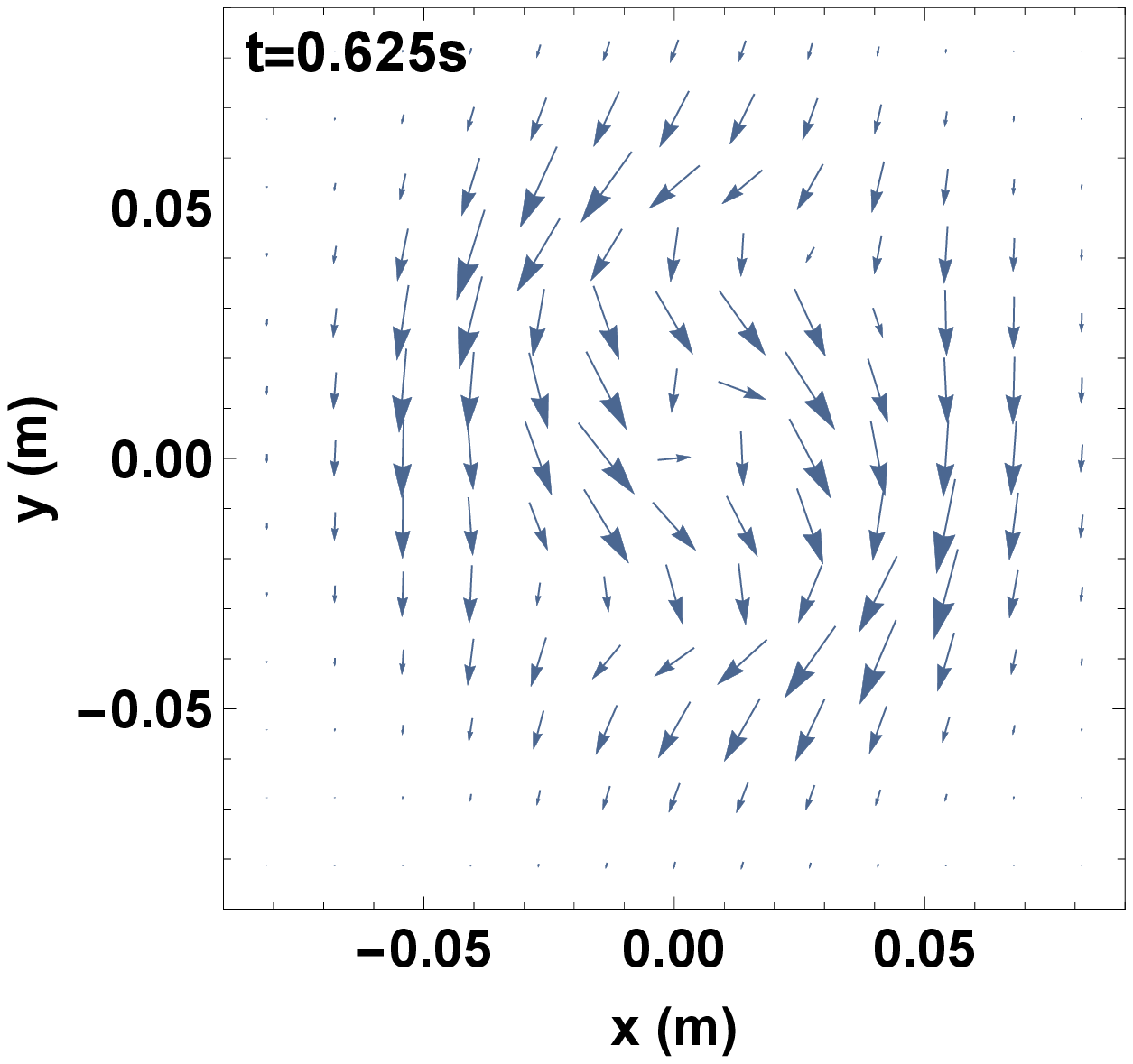}&\includegraphics[width=0.24\textwidth,angle=0]{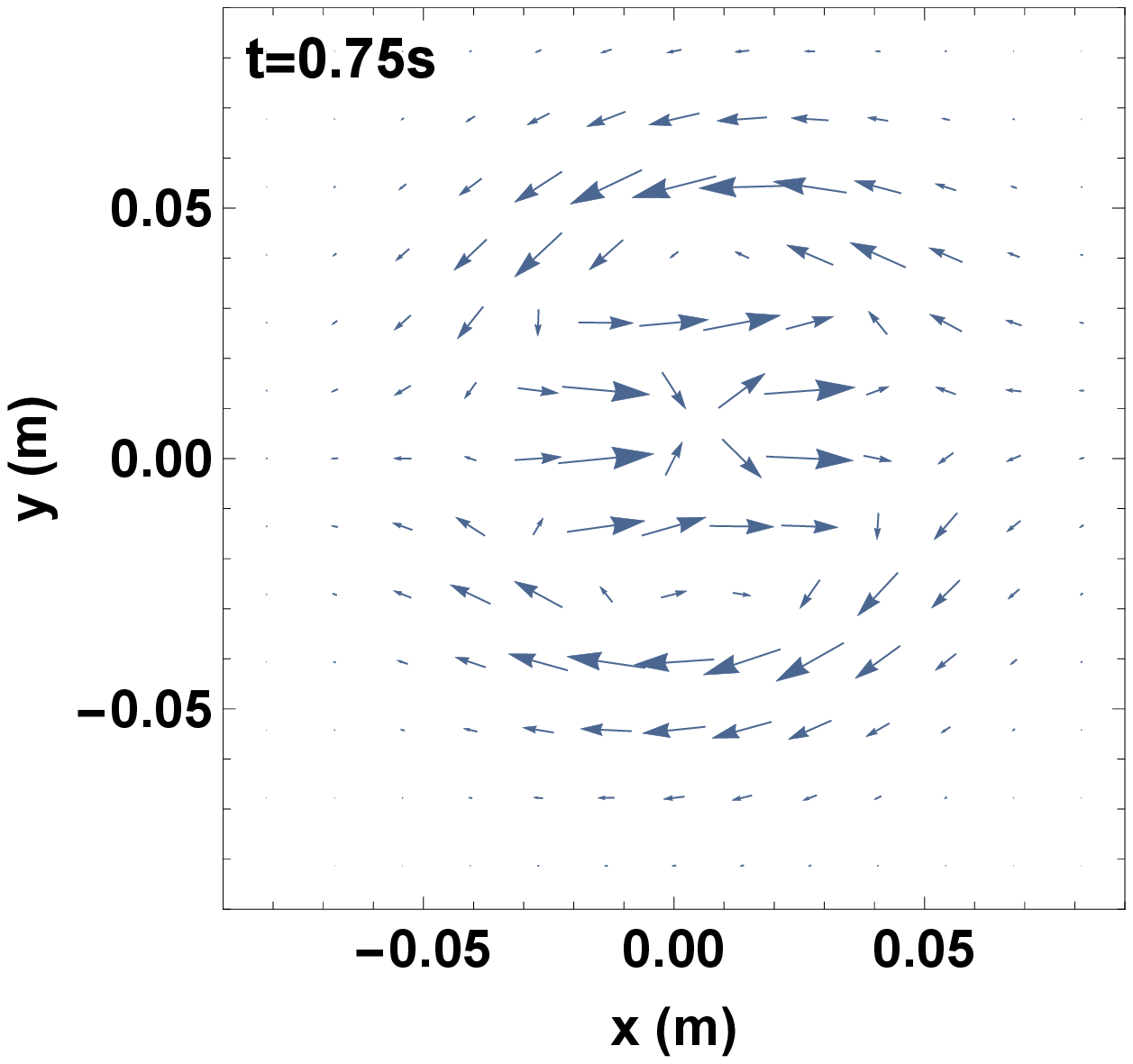}&\includegraphics[width=0.24\textwidth,angle=0]{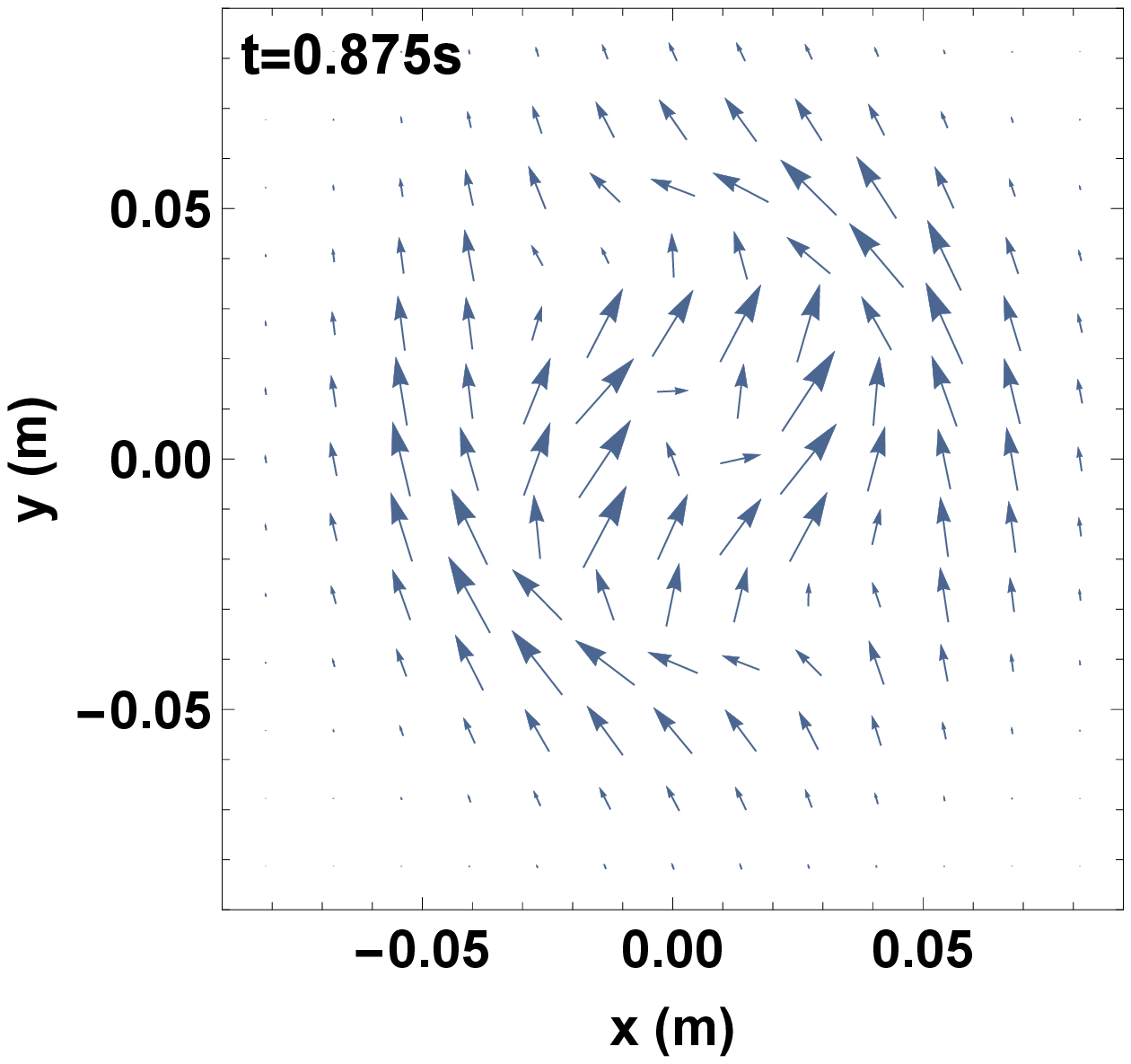}
\end{array}$
\end{center}\caption{Vector plots of perturbed mass flow for $m = 1$ mode in the cross section of MEVAT during one period of $1$~s.}
\label{fg6}
\end{figure*}

\subsection{Parameter dependence}\label{prm}
To guide the experimental design of an efficient MEVAT, this section is devoted to studying the effects of rotation frequency, field strength, electron temperature and cathode material on the growth rate and frequency of plasma instability. As shown in Fig.~\ref{fg7}, the normalized growth rate and frequency decrease when the rotation frequency is reduced (till $\Omega=0.2$), consistent with a previous observation,\cite{Chang:2011aa} because the centrifugal force is descending. However, when the rotation frequency drops below $\Omega=0.1$, the instability becomes much stronger and the profile of dispersion curve changes remarkably, indicating the entrance into a different fluctuation mode. 
\begin{figure}
\begin{center}$
\begin{array}{c}
\includegraphics[width=0.45\textwidth,angle=0]{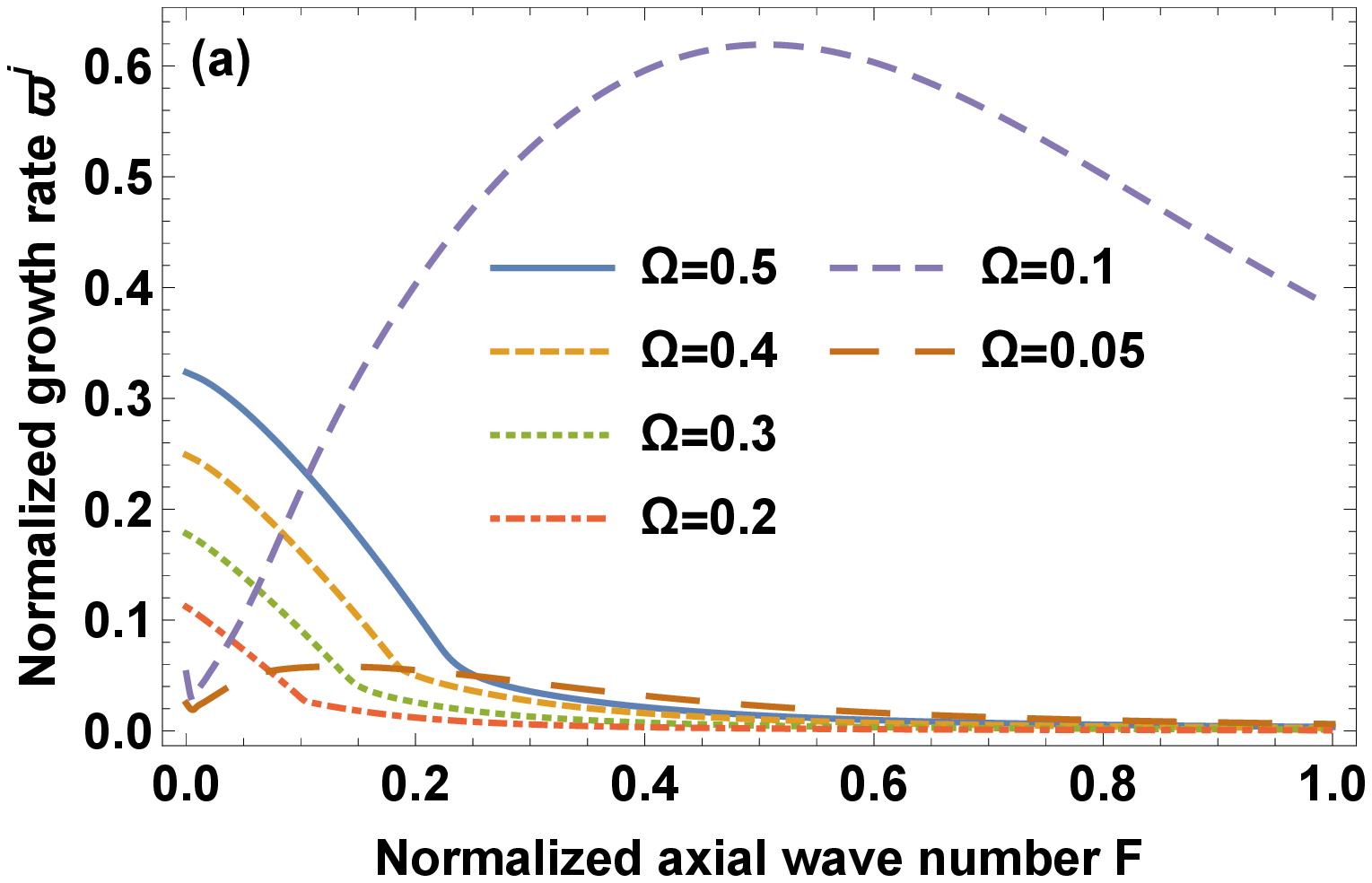}\\
\hspace{0.1 cm}\includegraphics[width=0.45\textwidth,angle=0]{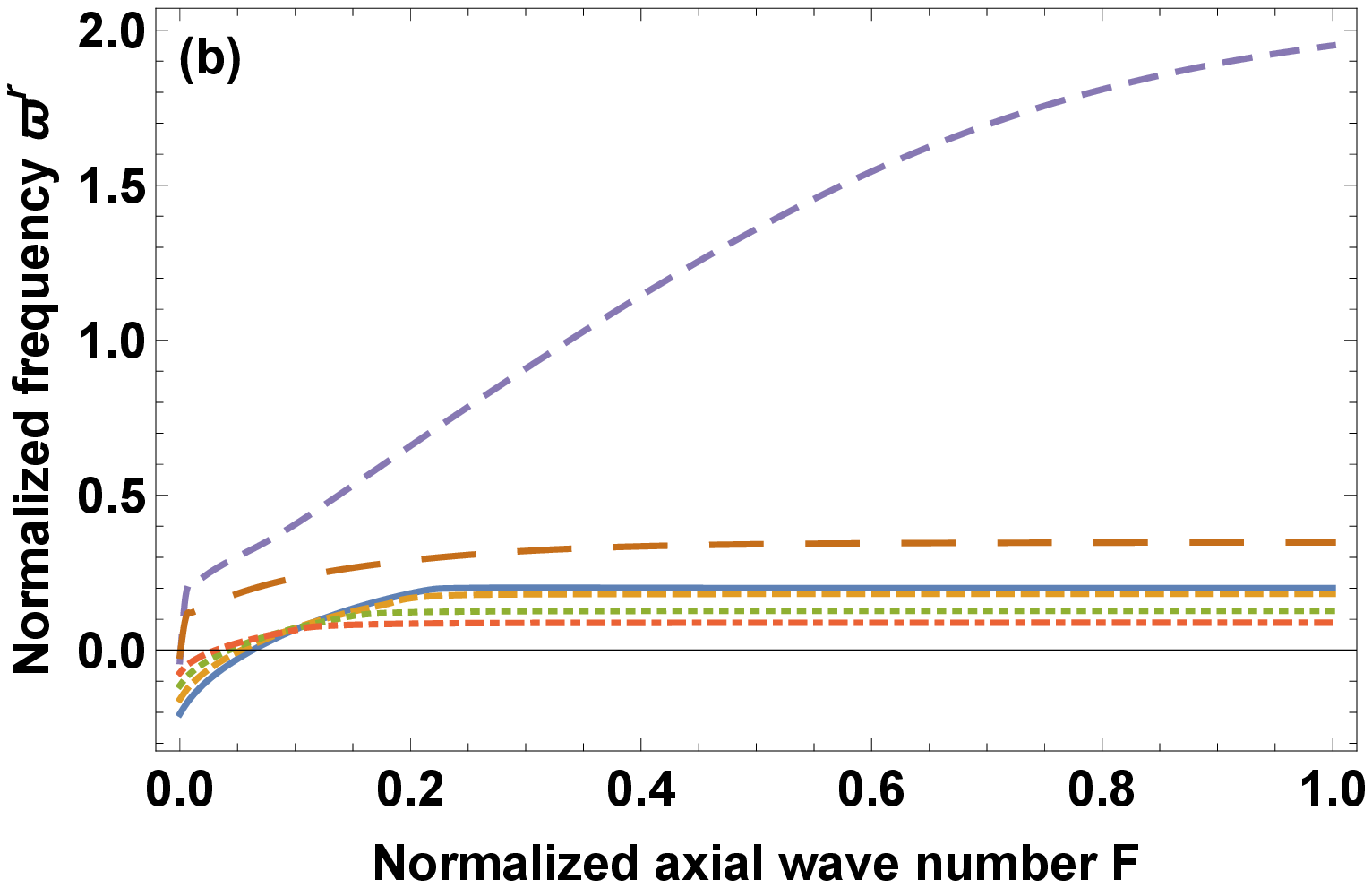}
\end{array}$
\end{center}\caption{Dependence of normalized growth rate $\varpi^i$ (a) and normalized frequency $\varpi^r$ (b) on plasma rotation frequency $\Omega$. (vs normalized axial wavenumber $F~=~k_\varsigma/\sqrt{\delta}$).}
\label{fg7}
\end{figure}
Figure~\ref{fg8} shows the influence of external magnetic field strength on the dispersion relation of MEVAT. Different from previous conclusion, however, the normalized growth rate increases with growing field strength, although the normalized frequency decreases as before.\cite{Chang:2011aa} 
\begin{figure}
\begin{center}$
\begin{array}{c}
\includegraphics[width=0.45\textwidth,angle=0]{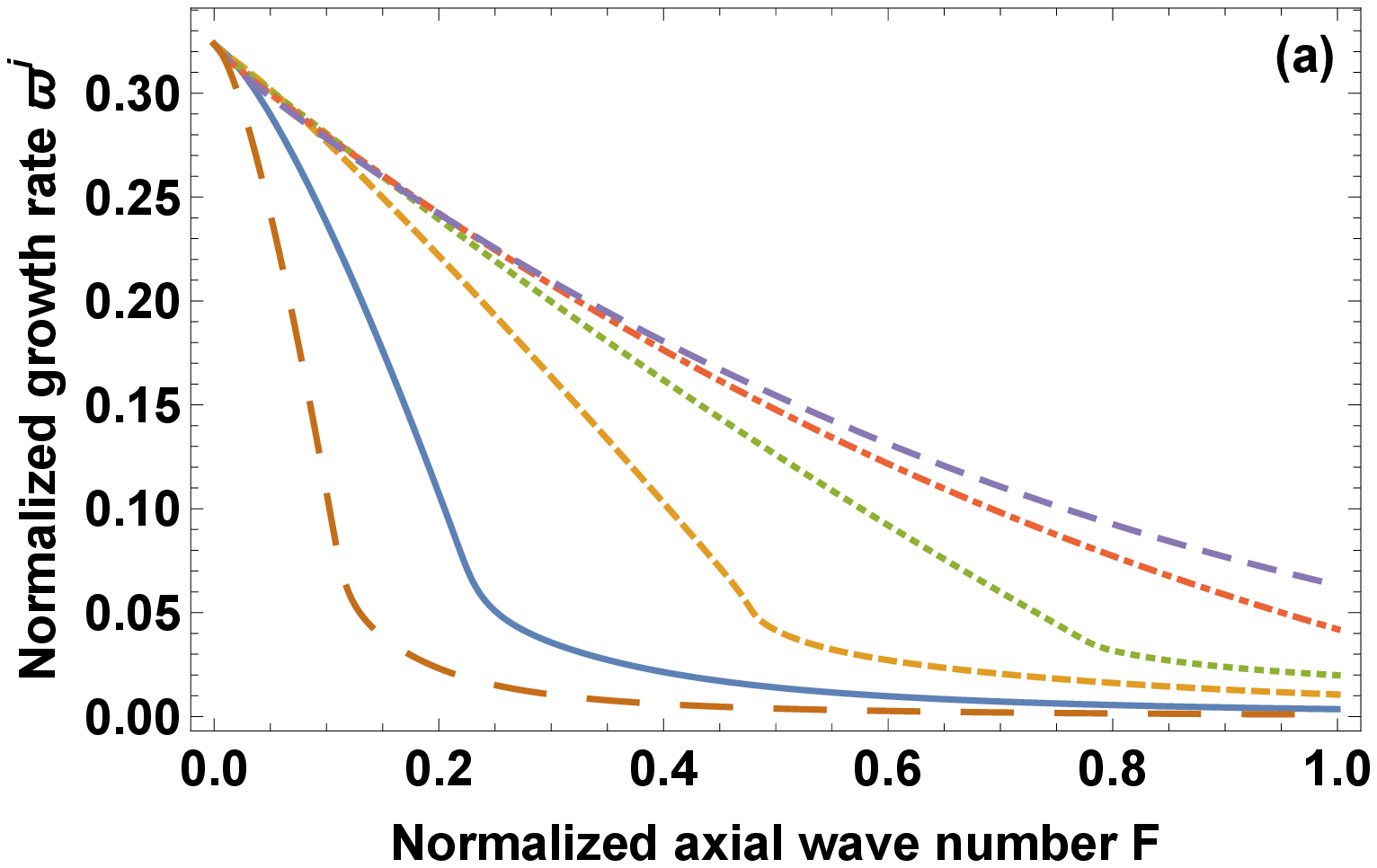}\\
\hspace{-0.18 cm}\includegraphics[width=0.46\textwidth,angle=0]{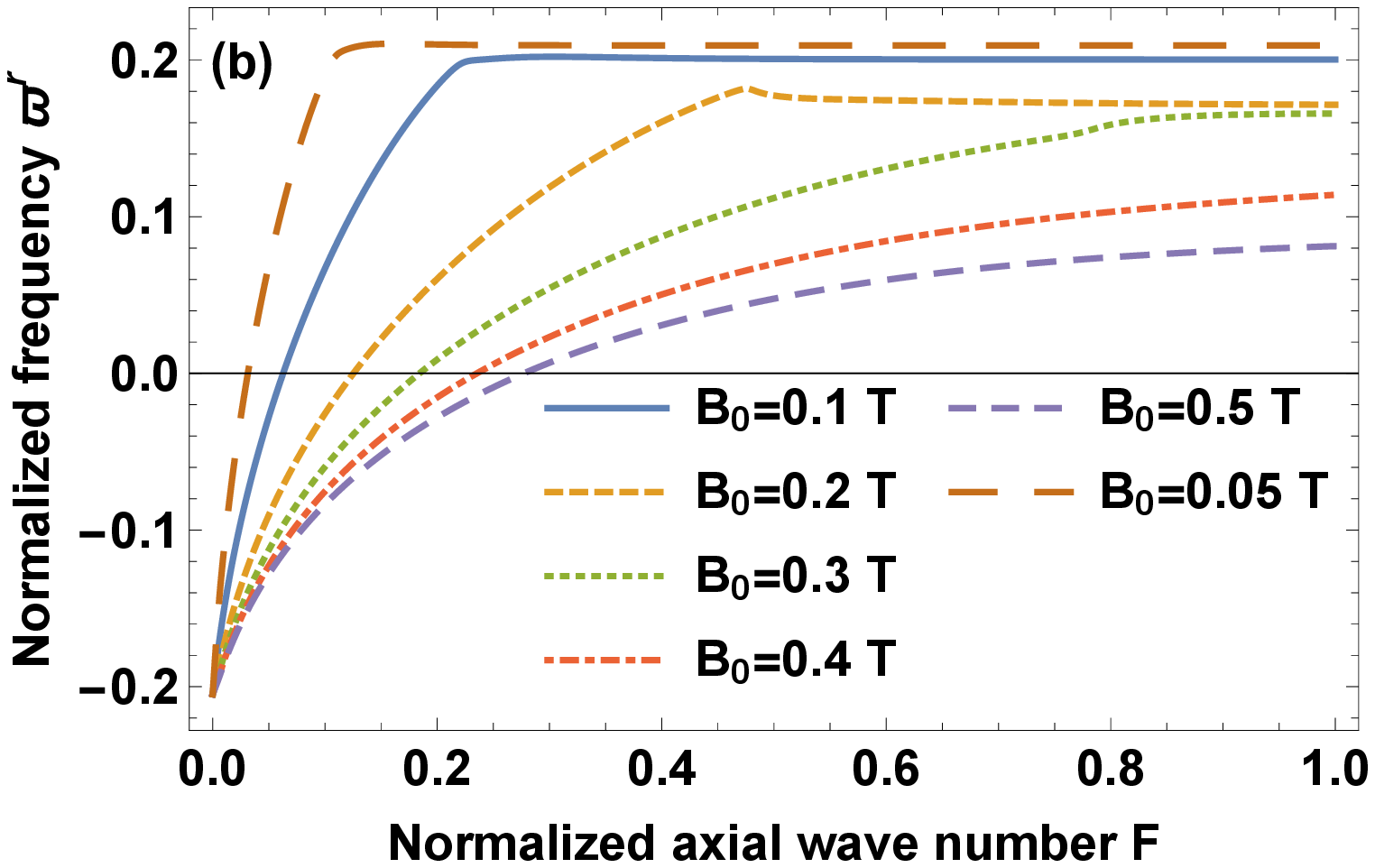}
\end{array}$
\end{center}\caption{Dependence of normalized growth rate $\varpi^i$ (a) and normalized frequency $\varpi^r$ (b) on external magnetic field strength $B_0$. (vs normalized axial wavenumber $F~=~k_\varsigma/\sqrt{\delta}$).}
\label{fg8}
\end{figure}
Moreover, as shown in Fig.~\ref{fg9}, the normalized growth rate also increases with descending electron temperature, different from previous observation, whereas the frequency drops slightly at expected.\cite{Chang:2011aa} We claim that this unusual variation of growth rate with field strength and electron temperature may be attributed to the extremely low normalized resistivity parallel to external field, namely $\delta=7.8\times 10^{-5}$, which makes the plasma nearly an ideal MHD medium. Additionally, we varied the ion temperature from $T_i=50$~eV to $T_i=10$~eV, and found that the change in dispersion curves is even smaller and negligible. 
\begin{figure}
\begin{center}$
\begin{array}{c}
\includegraphics[width=0.45\textwidth,angle=0]{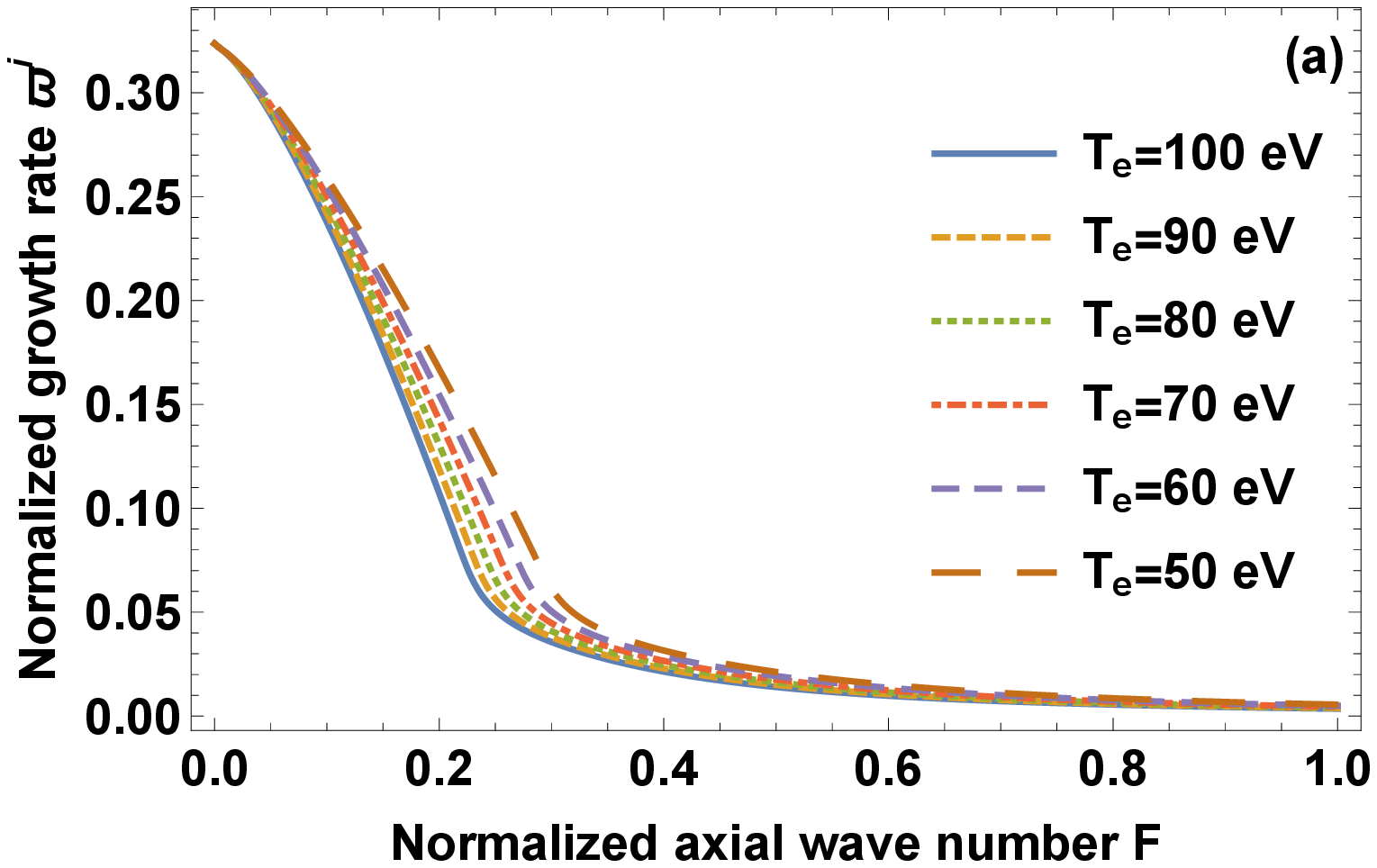}\\
\includegraphics[width=0.45\textwidth,angle=0]{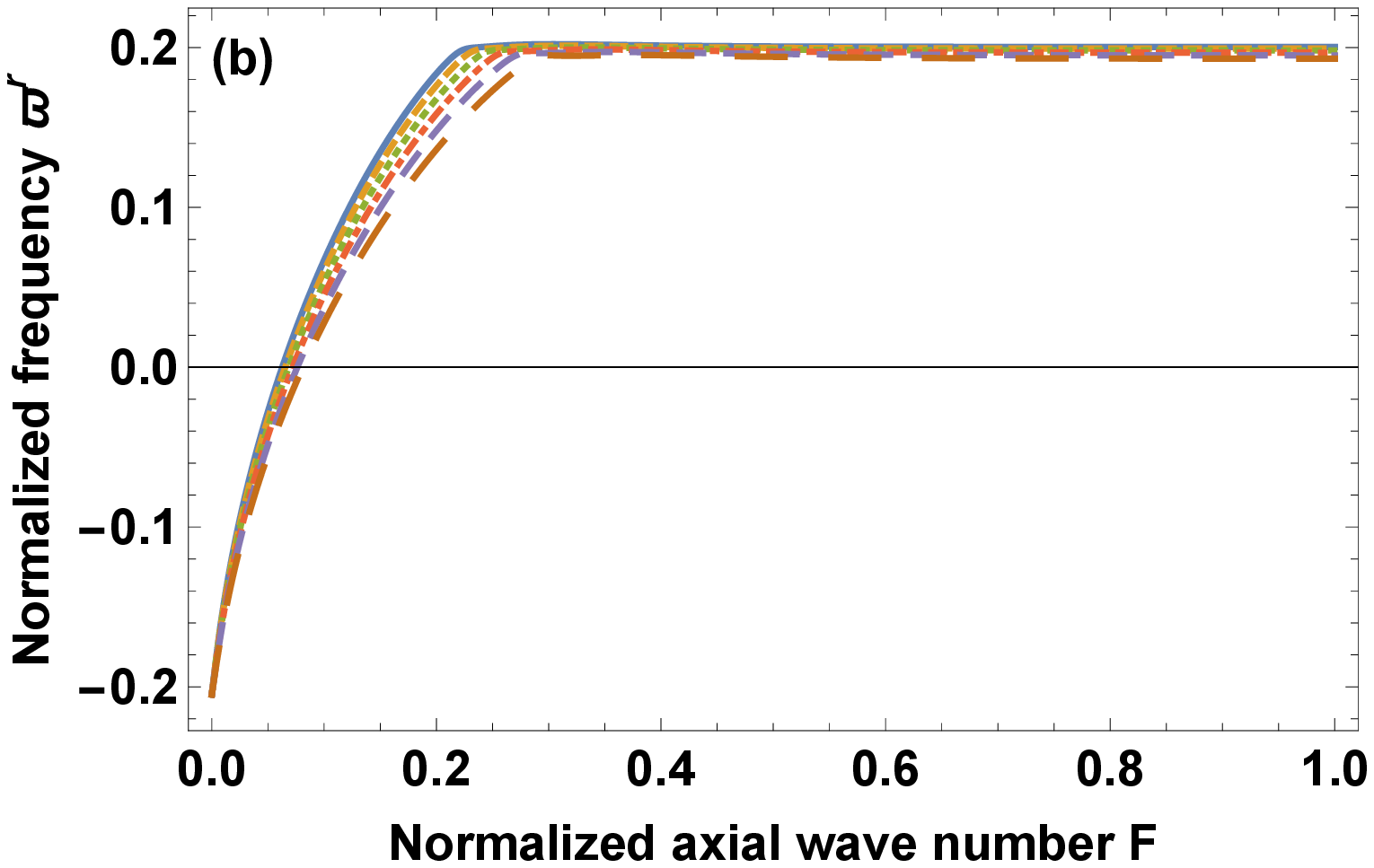}
\end{array}$
\end{center}\caption{Dependence of normalized growth rate $\varpi^i$ (a) and normalized frequency $\varpi^r$ (b) on electron temperature $T_e$. (vs normalized axial wavenumber $F~=~k_\varsigma/\sqrt{\delta}$).}
\label{fg9}
\end{figure}
Finally, we compare the plasma instability for different cathode materials, which are commonly used in MEVAT, including Ti for previous sections, magnesium (Mg), cuprum (Cu), molybdenum (Mo) and wolfram (W). As the atomic weight is increased, the normalized growth rate drops and the normalized frequency increases slightly, which is perhaps comprehensible in the sense that ions become heavier so that move slower. 
\begin{figure}
\begin{center}$
\begin{array}{c}
\includegraphics[width=0.45\textwidth,angle=0]{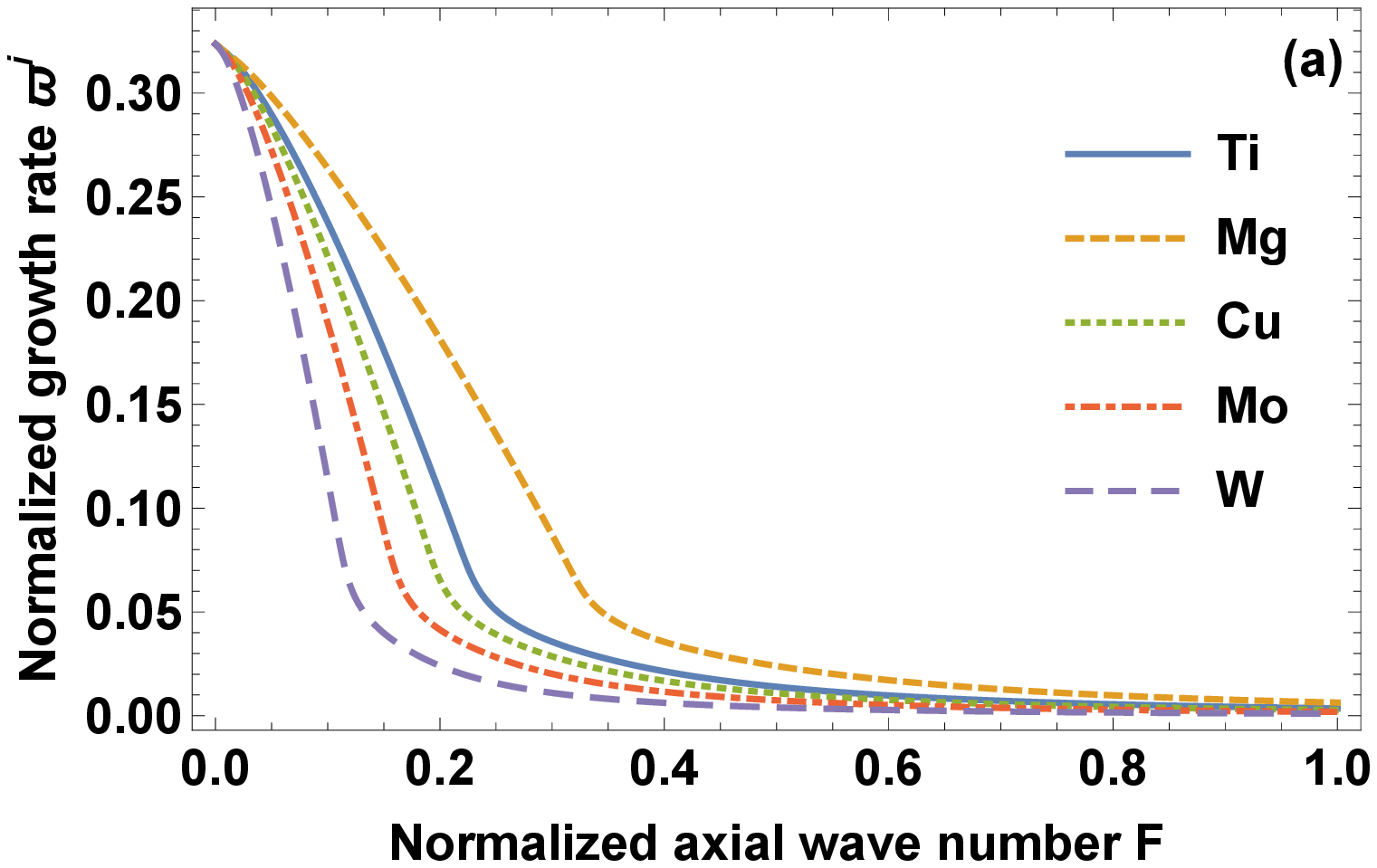}\\
\includegraphics[width=0.45\textwidth,angle=0]{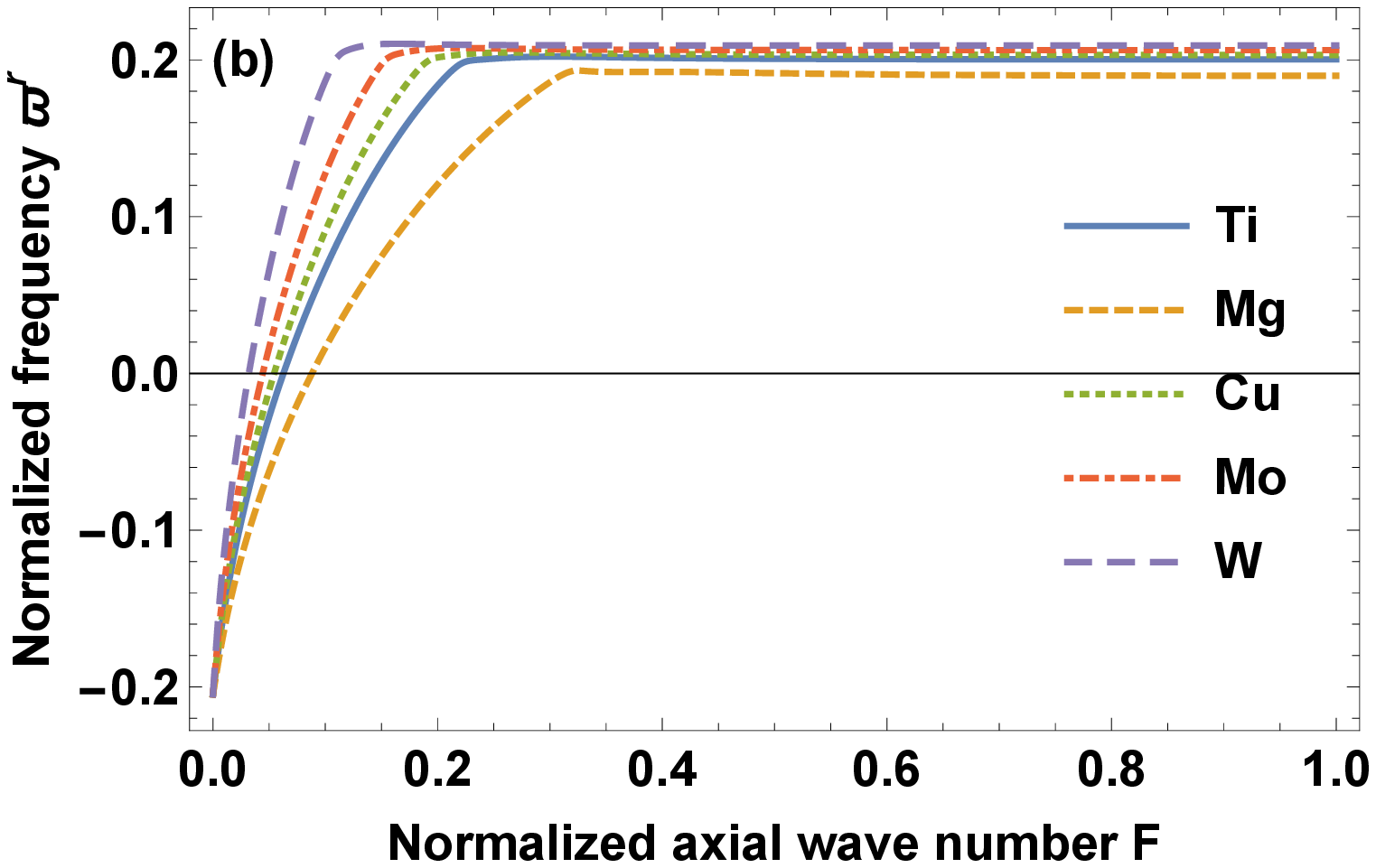}
\end{array}$
\end{center}\caption{Dependence of normalized growth rate $\varpi^i$ (a) and normalized frequency $\varpi^r$ (b) on cathode material. (vs normalized axial wavenumber $F~=~k_\varsigma/\sqrt{\delta}$).}
\label{fg10}
\end{figure}

\section{Conclusions}\label{cls}
To describe the plasma instability associated with plasma rotation and axial streaming, a two-fluid flowing plasma model developed originally for vacuum arc centrifuge is applied to MEVAT. Based on typical experimental data, the dispersion curve showing the growth rate and frequency of instability evolution is first computed via a numerical shooting method. The strongest instability occurs on axis for rotation frequency close to that of vacuum arc centrifuge, and its frequency is smaller than the sum of rotation frequency and axial velocity. Then the perturbed density is calculated through a linearized technique, which shows a radial mode transition from $n=0$ to $n=1$ at $r=4.24$~m, and it peaks near the radial location of maximum equilibrium density gradient, suggesting that the observed instability is a resistive drift mode driven by density gradient. Moreover, the temporal visualization of perturbed mass flow in the cross section of plasma column shows an anti-clockwise rotation (same to the direction of ion diamagnetic drift) inside the mode transition layer, and clockwise rotation (same to the direction of electron diamagnetic drift) outside. This implies that the plasma instability is dominated by different particle species in the core (ion) and at edge (electron). Finally, parameter scan shows that the instability strength increases with growing rotation frequency and field strength, and decreases with growing electron temperature and atomic weight. Possible reasons are suggested. Further research may focus on the comparison between these computations and experimental measurements on a well diagnosed MEVAT, and extend the analysis to other plasma thrusters where instability caused by plasma rotation and axial flow is of particular concern. These relevant thrusters may include helicon double layer thruster,\cite{Charles:2002aa} Hall thruster,\cite{Yu:2008aa} magnetized ion engine,\cite{Brophy:2002aa} radio-frequency plasma thruster,\cite{Lafleur:2011aa, Shabshelowitz:2013aa} micro-wave (electron cyclotron resonance) plasma thruster,\cite{Stallard:1996aa, Kuninaka:1998aa} and variable specific impulse magnetoplasma rocket (VASIMR),\cite{Chang-Diaz:2000aa} to name a few. 

\acknowledgements
This work is supported by various fundings: National Natural Science Foundation of China (11405271), China Postdoctoral Science Foundation (2017M612901), Chongqing Science and Technology Commission (cstc2017jcyjAX0047), Chongqing Postdoctoral Special Foundation (Xm2017109), Fundamental Research Funds for Central Universities (YJ201796), Pre-research of Key Laboratory Fund for Equipment (61422070306), and Laboratory of Advanced Space Propulsion (LabASP-2017-10). 

\section*{References}
\bibliographystyle{unsrt}

\end{document}